\documentclass[floatfix,prd,twocolumn,letterpaper,lengthcheck,superscriptaddress,showpacs,amssymb,amsmath,amsfonts,aps,altaffilletter,nofootinbib,nopreprintnumbers,showpacs,longbibliography]{revtex4-1}

\usepackage{color}
\usepackage{hyperref}
\usepackage{amsmath}
\usepackage{multirow}
\usepackage{graphicx}
\usepackage{caption}
\usepackage{subcaption}

\begin{document}

\title[]{Dynamics of marginally trapped surfaces in a binary black
  hole merger: Growth and approach to equilibrium}

\author{Anshu Gupta} 
\affiliation{Inter-University Center for Astronomy and Astrophysics,
  Post Bag 4, Ganeshkhind, Pune 411007, India}
\author{Badri Krishnan} 
\affiliation{Max-Planck-Institut f\"ur Gravitationsphysik (Albert
  Einstein Institute), Callinstr. 38, 30167 Hannover, Germany}
\author{Alex B. Nielsen}
\affiliation{Max-Planck-Institut f\"ur Gravitationsphysik (Albert
  Einstein Institute), Callinstr. 38, 30167 Hannover, Germany}
\author{Erik Schnetter}
\affiliation{Perimeter Institute for Theoretical Physics, Waterloo, 
  ON N2L 2Y5, Canada}
\affiliation{Department of Physics, University of Guelph, Guelph, 
  ON N1G 2W1, Canada}
\affiliation{Center for Computation \& Technology, Louisiana State
  University, Baton Rouge, LA 70803, USA}


\begin{abstract}

  The behavior of quasi-local black hole horizons in a binary black
  hole merger is studied numerically.  We compute the horizon
  multipole moments, fluxes and other quantities on black hole
  horizons throughout the merger.  These lead to a better qualitative
  and quantitative understanding of the coalescence of two black
  holes; how the final black hole is formed, initially grows and then
  settles down to a Kerr black hole.  We calculate the rate at which
  the final black hole approaches equilibrium in a fully
  non-perturbative situation and identify a time at which the linear
  ringdown phase begins.  Finally, we provide additional support for
  the conjecture that fields at the horizon are correlated with fields
  in the wave-zone by comparing the in-falling gravitational wave flux
  at the horizon to the outgoing flux as estimated from the
  gravitational waveform.

\end{abstract}

\maketitle

\section{Introduction}
\label{sec:intro}

Gravitational wave signals from binary black hole merger events are
now routinely computed in numerical simulations.  For generic spin
configurations and at least for moderate mass ratios, various aspects
of the problem are well understood; this includes the initial data,
numerical methods, gauge conditions for the evolution, locating
black hole horizons, and finally extracting gravitational wave signals
in the wave zone.  We refer the reader to
\cite{Pretorius:2005gq,Campanelli:2005dd,Baker:2005vv} which
demonstrated the first successful binary black hole simulations and to
e.g. \cite{BaumgarteShapiroBook,ShibataBook,AlcubierreBook} for
further details and references.

What is somewhat less well understood is the behavior of black hole
horizons near the merger.  For example, it is clear from the black
hole area increase law that as soon as the final black hole is formed,
its area will increase and it will eventually asymptote to a higher
value. It is also expected that the rate of area increase will be
largest immediately when the common horizon is formed which is when
the in-falling gravitational radiation has the largest amplitude and
the effects of non-linearities cannot be neglected.  At a somewhat
later time, the rate of area increase will slow down and the problem
can be treated within black hole perturbation theory.  Unanswered
questions include: What is the angular distribution of the
gravitational wave flux entering the horizon and causing it to grow?
What is the rate of decrease of this flux with time?  Is it possible
to identify a time, purely based on the properties of the horizon,
after which the flux is small enough that we can trust the results of
perturbation theory?

A different but related set of questions arises with the approach to
Kerr.  Mass and spin multipole moments of black hole horizons can be
constructed \cite{Ashtekar:2004gp,Ashtekar:2013qta}.  These multipole
moments describe the instantaneous intrinsic geometry of the black
hole at any given time.  The first mass multipole moment is just the
mass, while the first non-zero spin multipole moment for a regular
horizon is just the angular momentum.  For a Kerr black hole, these
lowest multipole moments determine uniquely all of the higher moments.
For the dynamical black hole, the higher moments can be computed
independently in the numerical simulation and we can extract the rate
at which the multipole moments approach their Kerr values. Some of the
above questions were considered in \cite{Schnetter:2006yt} but
numerical relativity has made great progress since then and it is
useful to revisit these issues again.  We shall use the framework of
quasi-local horizons for our analysis; see
e.g. \cite{Ashtekar:2004cn,Booth:2005qc} for reviews.

Another question we wish to address is that of when the gravitational
waveform can be considered to be in the ringdown phase.  In principle,
by fitting the final part of the waveform with damped sinusoids we can
extract the frequency $f$ and damping time $\tau$ of the black hole
ringdown mode(s), thereby allowing a test of the Kerr nature of the
final black hole proposed in \cite{Dreyer:2003bv} if we can observe
more than a single mode.  However, it is non-trivial to know at what
point one should start fitting the damped sinusoid.  If we start too
close to the merger, incorrect values of $(f,\tau)$ can be obtained.
An example of this issue appears in the ringdown analysis of the
binary black hole detection GW150914 \cite{TheLIGOScientific:2016src}.
As demonstrated in Fig. 5 of \cite{TheLIGOScientific:2016src},
choosing different start-times for fitting a damped sinusoid to the
post-merger phase of the observed strain data has a noticeable effect
on the recovered values of the frequency and damping time, and thus
also on the inferred values of the mass and angular momentum of the
final black hole.  Similar questions have been studied recently in
\cite{Thrane:2017lqn,Bhagwat:2017tkm,Cabero:2017avf}. It is then
natural to ask whether one can use correlations with the horizon to
quantitatively provide a time in the waveform beyond which the
ringdown analysis is valid.  We shall use the multipole moments to
address this question.  

Finally, for vacuum general relativity, the behavior of spacetime at
or near the horizon is \emph{correlated} with what happens in the wave
zone and the gravitational waveform.  This notion was discussed in a
series of papers by Jaramillo et al.
\cite{Jaramillo:2011rf,Jaramillo:2011re,Jaramillo:2012rr}\footnote{It is
  likely that related ideas were also motivating factors for earlier
  work, such as the so-called stretched horizon in the membrane
  paradigm \cite{Price:1986yy}; but the notion of correlations is not
  mentioned explicitly.} and used to
explain the phenomenon of {\it{anti-kicks}}, a short phase of
deceleration which reduces the kick-velocity of the final black hole
remnant \cite{Rezzolla:2010df}.  The basic idea is quite simple: in an
initial value formulation of general relativity, the initial data
fields at early times outside the horizon determine both the behavior
of the horizon and also the waveform far away from the horizon.  Thus,
if the in- and out-going modes are coupled (as they most likely are
due to the non-linearities of general relativity), there must be
correlations between data on the horizon and the gravitational
waveform.  This applies also to fields on apparent horizons which are
inside the event horizon.  As expected the black hole horizon is not a
source for the gravitational waveform and of course cannot causally
influence any observations in the wave zone.  However, correlations
due to a common source could provide a way to extract information
about the near horizon spacetime from gravitational wave observations.
A similar suggestion was also made in Section 8 of
\cite{Ashtekar:2004cn}: the radiation trapped between the horizon and
the peak of the effective gravitational potential outside the black
hole could fall into the horizon thereby increasing its area, and also
cause the black hole horizon to lose its irregularities.  This apparently simple conjecture is not
yet fully developed. For example, it could be possible that such
correlations do not exist for generic initial data but are instead a
special property of astrophysical initial data where one wants to
minimize incoming radiation from past null infinity.  Here we shall
provide additional support to this conjecture by showing that the
outgoing flux obtained from the gravitational signal is highly
correlated with the in-falling flux at the horizon.

The plan for the rest of the paper is as follows.  Section
\ref{sec:prelim} briefly reviews basic notions and equations for
dynamical horizons and the numerical simulations.  Section
\ref{sec:finalbh} describes various dynamical horizon quantities
computed numerically in a binary black hole simulation providing a
better qualitative and quantitative understanding of binary black hole
coalescence.  This section also illustrates the area increase law
starting with the initial black hole and ending with the final black
hole. It is unclear whether or not there is a connected sequence of
marginally trapped surfaces that take us from the initial black holes
to the final one. If in fact there is such a sequence of marginally
trapped surfaces, we could track the area of the black hole through
the merger.  Sec.~\ref{sec:multipoles} studies the horizon multipole
moments and the rate at which they approach their equilibrium Kerr
values.  We identify an epoch about $10M$ after the formation of the
common horizon when there is an evident change in the decay rate of
the moments.  Section \ref{sec:cross-corr} carries out the
cross-correlation study between the horizon fluxes with the waveform
(or more precisely, with the outgoing luminosity).  This provides the
critical link between properties of spacetime in the strong field
region and gravitational wave observations.  Section
\ref{sec:conclusions} presents concluding remarks and lists some open
problems.

\section{Preliminaries}
\label{sec:prelim}

\subsection{Basic properties of dynamical horizons}
\label{subsec:dh}

We begin by briefly summarizing basic definitions of marginally
trapped surfaces and dynamical horizons for later use.  Let
$\mathcal{S}$ be a closed spacelike surface with topology
$S^2$. Denote the out- and in-going future directed null normals to
$\mathcal{S}$ by $\ell^a$ and $n^a$ respectively\footnote{We shall use
  the abstract index notation with $g_{ab}$ denoting the spacetime
  metric of signature $(-,+,+,+)$, $\nabla_a$ the derivative operator
  compatible with $g_{ab}$, and the Riemann tensor defined as
  $2\nabla_{[a}\nabla_{b]}X_c = {R_{abc}}^dX_d$ for any 1-form
  $X_c$.}.  We require that $\ell\cdot n = -1$.  We are allowed to
scale $\ell^a$ and $n^a$ by positive definite functions $f$ such that
$\ell^a \rightarrow f\ell^a$ and $n^a \rightarrow f^{-1}n^a$, thus
preserving $\ell\cdot n$.  Let $q_{ab}$ be the intrinsic 2-metric on
$\mathcal{S}$ obtained by restricting the spacetime metric to
$\mathcal{S}$.  The expansions of $\ell^a$ and $n^a$ are defined as
\begin{equation}
  \Theta_{(\ell)} = q^{ab}\nabla_a\ell_b\,,\qquad \Theta_{(n)} = q^{ab}\nabla_an_b\,.  
\end{equation}
$\mathcal{S}$ is said to be a marginally outer trapped surface (MOTS)
if $\Theta_{(\ell)} = 0$.  In addition, $\mathcal{S}$ is said to be a
(future-outer) marginally trapped surface if $\Theta_{(n)}< 0$.  MOTSs
are used to conveniently locate black holes in numerical simulations
\cite{Thornburg:2003sf}.  This can be done on every time slice, and
does not require knowledge of the full spacetime.

Under certain general stability conditions, it can be shown that
marginally trapped surfaces evolve smoothly under time evolution
\cite{Andersson:2005gq,Andersson:2008up,Andersson:2007fh}.  Apparent
horizons are the outermost marginally outer trapped surfaces on a
given spatial slice and the outermost condition can cause apparent
horizons to jump discontinuously as new marginally trapped surfaces
are formed.  However, the underlying marginally trapped surfaces
continue to evolve smoothly.  We shall describe this behavior in
greater detail later in this paper. Even the marginally trapped
surfaces which are not expected to satisfy the stability conditions
are found empirically to evolve smoothly thus a more general result
might hold for the time evolution \cite{Booth:2017fob}.

Given the smooth time evolution, we can consider the sequence
$\mathcal{S}_t$ of marginally outer trapped surfaces at various times
$t$ and construct the 3-dimensional tube obtained by stacking up all
the $\mathcal{S}_t$.  This leads us to the definition of a marginally
trapped tube (MTT) as a 3-surface foliated by marginally outer trapped
surfaces.  Let $\mathcal{H}$ denote the MTT, and let $A_t$ be the area
of $\mathcal{S}_t$.  There are three cases of interest depending on
whether $\mathcal{H}$ is spacelike, null or timelike.  When
$\mathcal{H}$ is spacelike, and $\Theta_{(n)} < 0$, it is called a
dynamical horizon
\cite{Ashtekar:2002ag,Ashtekar:2003hk,Ashtekar:2004cn}.  In this case,
its area increases monotonically in the outward direction.  More
generally, this also holds if the average of $\Theta_{(n)}$ over
$\mathcal{S}_t$ is negative.  If $\widehat{r}^a$ is the unit normal
(outward pointing) to $\mathcal{S}_t$ on $\mathcal{H}$, and
$\widehat{\tau}^a$ is the future directed unit timelike normal to
$\mathcal{H}$, we can construct the out- and in-going null normals
\begin{equation}
  \label{eq:dhnullnormals}
  \ell^a = \frac{\widehat{\tau}^a+\widehat{r}^a}{\sqrt{2}}\,,\quad n^a = \frac{\widehat{\tau}^a-\widehat{r}^a}{\sqrt{2}}\,.
\end{equation}
It then follows that
\begin{equation}
  \mathcal{L}_{\widehat{r}} \widetilde{\epsilon}_{ab} = -\frac{1}{\sqrt{2}}\Theta_{(n)}\widetilde{\epsilon}_{ab}\,,
\end{equation}
where $\widetilde{\epsilon}_{ab}$ is the volume 2-form on
$\mathcal{S}_t$.  Integrating this equation over $\mathcal{S}_t$ shows
that the area increases along $\widehat{r}^a$ if the average of
$\Theta_{(n)}$ on $\mathcal{S}_t$ is negative.  

When $\mathcal{H}$ is null the areas of its cross-sections
$\mathcal{S}_t$ are constant, and is known as an isolated horizon (see
e.g. \cite{Ashtekar:2000sz,Ashtekar:1998sp,Ashtekar:1999yj,Ashtekar:2001is,Ashtekar:2001jb}
for details and precise definitions).  When $\mathcal{H}$ is timelike
and $\Theta_{(n)}<0$, the area of its cross-sections decreases to the
future and is called a timelike membrane.  Timelike MTTs appear in
numerical simulations as well (but they are not the outermost
marginally outer trapped surfaces).  The reader is referred to
\cite{Ashtekar:2004cn,Booth:2005qc,Krishnan:2013saa} for reviews. See
e.g. \cite{BenDov:2004gh,Booth:2005ng} for examples of MTTs of various
signatures; see also \cite{Chu:2010yu} for other numerical studies of
inner and outer horizons.

For a dynamical horizon, we can in fact do better than just showing
that the area increases. As at null infinity, we can obtain an
explicit positive-definite expression for the flux of gravitational
radiation crossing a dynamical horizon.  This is a non-trivial fact
which does not hold for arbitrary surfaces, and it emphasizes again
the special properties of a dynamical horizon
\cite{Ashtekar:2002ag,Ashtekar:2003hk,Ashtekar:2004cn}.  To discuss
this further we need a few more definitions.  The shear $\sigma_{ab}$
of $\ell^a$ will play an important role and it is defined as:
\begin{equation}
  \label{eq:shear-defn}
  \sigma_{ab} = {q_a}^c{q_b}^d\nabla_c\ell_d - \frac{1}{2}\Theta_{(\ell)}q_{ab}\,.
\end{equation}
Let $h_{ab}$ be the 3-metric on $\mathcal{H}$ and as before let
$\widehat{r}^a$ be the outward pointing unit spacelike normal to
$\mathcal{S}_t$ on $\mathcal{H}$.  Define a one-form $\zeta_a$ on
$\mathcal{H}$ as $\zeta_a = {h_a}^b\widehat{r}^c\nabla_c\ell_b$.  The
instantaneous gravitational energy flux is an integral of a quantity
$\mathfrak{f}$ over marginally trapped surfaces $\mathcal{S}_t$:
\begin{equation}
  \label{eq:dhflux}
  \mathfrak{f} = \frac{1}{2}\sigma_{ab}\sigma^{ab} + \zeta_a\zeta^a \,.
\end{equation}
$\mathfrak{f}$ is the energy flux per unit area and per unit time
entering the horizon (with the area radius of $\mathcal{S}_t$ playing
the role of ``time'' on the horizon); note the different normalization
of $\ell\cdot n$ compared to \cite{Ashtekar:2004cn}.  This is an exact
expression in full general relativity with no approximations. It
satisfies the expected properties of gravitational radiation, for
example it is manifestly positive and it vanishes in spherical
symmetry.  No such local expression is possible for the event horizon
in general.  This is because of the global properties of the event
horizon; there are well known examples where the event horizon grows
in flat space where there cannot be any non-zero local flux
\cite{Ashtekar:2004cn}.  There do exist flux formulae for the growth
of the event horizon in perturbative situations. In
\cite{Hawking:1972hy} it is found that the rate of area increase for
an event horizon is approximately proportional to the integral of
$|\sigma|^2$ with $\sigma$ being the shear of the null generator of
the event horizon.  This however only holds within perturbation theory
and furthermore, because of the nature of the event horizon, this
really only makes sense when the end state of the event horizon is
known or assumed.  See \cite{Booth:2006bn} for a more detailed
comparison with \cite{Hawking:1972hy}.

The other ingredient we shall use frequently in this paper are the
multipole moments.  These were first introduced in
\cite{Ashtekar:2004gp} for isolated horizons, and extended and used in
\cite{Schnetter:2006yt} for dynamical horizons.  These multipole
moments have found applications in, for example, predictions of the
anti-kick in binary black hole mergers \cite{Rezzolla:2010df} and for
studying tidal deformations of black holes
\cite{Cabero:2014nza,Gurlebeck:2015xpa}.  The work by Ashtekar et al.
\cite{Ashtekar:2013qta} provides flux formulae for the multipole
moments and a procedure for choosing a suitable class of time
evolution vector fields on a dynamical horizon.  Here we shall use
them to study the approach of a dynamical horizon to equilibrium.

Our investigation of dynamical horizons will be informed by exact
results for axisymmetric isolated horizons $\mathcal{H}$.  Every
cross-section of an isolated horizon with spherical topology has the
same area $A$.  Let $\ell^a$ be a null generator of $\mathcal{H}$ and
$\varphi^a$ the axial symmetry vector field. For an isolated horizon
it can be shown that the Weyl tensor component $\Psi_2$ at the horizon
is time independent.  On every cross-section $\mathcal{S}$ it is given
by
\begin{equation}
  \Psi_2 = \frac{\mathcal{R}}{4} + \frac{i}{2} {}^\star d\omega\,.
\end{equation}
Here $\mathcal{R}$ is the two-dimensional scalar curvature of
$\mathcal{S}$, ${}^\star$ denotes the Hodge-dual, and $\omega$ is a
1-form on $\mathcal{H}$ defined by
\begin{equation}
  V^a\nabla_a \ell_b =  V^a\omega_a \ell_b\,
\end{equation}
with $V^a$ being any vector field tangent to $\mathcal{H}$.  The
surface gravity is $\kappa_{\ell} := \omega_a\ell^a$. The angular
momentum of $\mathcal{H}$ is given by 
\begin{equation}
  J_{\mathcal{S}} = -\frac{1}{8\pi}\oint_{\mathcal{S}} \omega_a\varphi^a \,d^2S\,.
\end{equation}
Let $\Sigma$ be a spatial Cauchy surface which intersects
$\mathcal{H}$, and let $\mathcal{S} = \mathcal{H}\cap \Sigma$.  It
turns out that $\omega_a\varphi^a = K_{ab}\widehat{R}^a\varphi^a$
where $K_{ab}$ is the extrinsic curvature of $\Sigma$ and $R^a$ the
unit spacelike normal to $\mathcal{S}$ on $\Sigma$
\cite{Dreyer:2002mx}. We will use this exact result for isolated
horizons to also define angular momentum for cross-sections of
dynamical horizons, since $K_{ab}$ is readily available in any
numerical simulation based on a 3+1 formulation of general relativity.
Given the angular momentum, the horizon mass is then defined by
\begin{equation}
  M_{\mathcal{S}} = \frac{1}{2R_{\mathcal{S}}}\sqrt{R_{\mathcal{S}}^4 + 4J_{\mathcal{S}}^2}\,,
\end{equation}
with $R_{\mathcal{S}}=\sqrt{A/4\pi}$ being the area-radius of $\mathcal{S}$.

The symmetry vector $\varphi^a$ can be used to construct a preferred
coordinate system $(\theta,\varphi)$ on $\mathcal{S}$ analogous to the
usual spherical coordinates on a sphere.  We can then use spherical
harmonics in this preferred coordinate system to construct multipole
moments.  As expected we have two sets of moments $M_n$, $J_n$ such
that $M_0$ is the mass $M_{\mathcal{S}}$ and $J_1$ is the angular
momentum $J_{\mathcal{S}}$.  Moreover, $\mathcal{R}$ and $\omega_a$
can be thought of as being (proportional to) the surface mass density
and surface current on $\mathcal{S}$ respectively
\cite{Ashtekar:2004nd}.  This leads us to the following expressions
for the multipole moments:
\begin{equation}
  M_n = \frac{M_{\mathcal{S}}R_{\mathcal{S}}^n}{8\pi}\oint_S \mathcal{R} P_n(\zeta) d^2S\,,
\end{equation}
and 
\begin{equation}
  J_n = \frac{R_{\mathcal{S}}^{n-1}}{8\pi} \oint_S P_n^\prime(\zeta) K_{ab}\varphi^a R^b d^2S \,.
\end{equation}
Here $P_n(\zeta)$ is the $n^{\mathrm{th}}$ Legendre polynomial, $P_n^\prime(\zeta)$ its derivative and
$\zeta = \cos\theta$.  

For a dynamical horizon, many of the above assumptions do not hold.
For example, $\Psi_2$ is not time independent and neither are the
area, curvature other geometric quantities on $\mathcal{S}_t$.
However, following \cite{Schnetter:2006yt}, we shall continue to
interpret the surface density and current in the same way so that the
multipole moments share the same definitions as above. These multipole
moments are gauge independent in the same sense as a dynamical horizon
is gauge independent, i.e. it exists as a geometric object in spacetime
independent of the spacetime foliation used to locate it.  A different
choice of spacetime slicing will give a different dynamical horizon,
but for any given dynamical horizon, the multipole moments are gauge
independent. 

An important issue is the choice of $\varphi^a$.  When the common
horizon is formed, it is highly distorted and it will generally not be
even approximately axisymmetric.  For this reason, while there exist
various methods for finding approximate axial Killing vectors
\cite{Cook:2007wr,Owen:2007dya,Lovelace:2008tw,Beetle:2008yt,Dreyer:2002mx,Beetle:2013zga},
in our opinion it is not fruitful to try and apply these to the newly
formed common horizon such as the ones we have here.  Since we shall
restrict ourselves to the case of the merger of equal mass
non-spinning black holes where the orbital angular momentum provides a
natural orientation, the kick velocity for the final black hole
vanishes, and the spacetime has reflection symmetry.  This initial
configuration is physically relevant since all gravitational wave
events observed from binary black hole mergers so far are consistent
with being comparable mass systems of initially non-spinning black
holes.  We shall therefore align the z-axis with the orbital angular
momentum and simply take $\varphi^a$ to be $\partial_\phi$,
i.e. defined by the z-axis.  The presence of reflection symmetry
across the equator makes this a natural (though of course not unique)
choice as well.  For more general initial configurations, we expect
the approach suggested in \cite{Ashtekar:2013qta} to be useful since
it relies only on the end-state being axisymmetric with an axial
symmetry vector $\varphi^a$ and it provides a method of transporting
$\varphi^a$ to all points on the dynamical horizon.  This will be
implemented in forthcoming work.

\subsection{Numerical simulations of binary black hole mergers}
\label{subsec:nr}

\subsubsection{Physical setup}

We employ a full numerical simulation to generate a binary black hole
spacetime geometry. As we are interested only in the merger and
ringdown phases of a binary black hole merger, we start our simulation
shortly before the merger, choosing the so-called QC-0 initial
conditions \cite{Baker:2002qf, Cook:1994va} for simplicity. These
correspond to an equal-mass non-spinning binary black hole system in
its last orbit before coalescence.

The QC-0 system has an ADM mass $M_{\mathrm{ADM}} \approx 1.00788\,M$,
where $M$ is the (arbitrarily chosen) mass unit in the
simulation. Compared to calculations that track several orbits of the
inspiral phase, the main difference of our setup is that it does not
give us access to the inspiral waveform, and that we do not know the
eccentricity that the QC-0 would have had during inspiral.  We list
the QC-0 system parameters in table \ref{tab:qc0}.

\begin{table}
  \begin{tabular}{lll}
    Parameter & Symbol & Value \\
    \hline
    half separation & $b$ & $1.168642873$ \\
    puncture mass & $m^+$ & $0.453$ \\
    puncture mass & $m^-$ & $0.453$ \\
    puncture momentum & $p^+_y$ & $+0.3331917498$ \\
    puncture momentum & $p^-_y$ & $-0.3331917498$ \\
    \hline
    total mass & $M_{\mathrm{ADM}}$ & $1.00788$ \\
  \end{tabular}
  \caption{QC-0 system parameters}
  \label{tab:qc0}
\end{table}

We track the two individual apparent horizons, and we find that the
system performs about three quarters of an orbit before a common
apparent horizon forms. We locate both the outer and the inner common
apparent horizons (see e.g. figure \ref{fig:area}), and by comparing the
shapes and areas of the common horizons we verify that the common
horizons form a single smooth world tube, and that we detect this
common horizon immediately as it appears in our spacetime
foliation. After coalescence, the outer common horizon quickly settles
down to a stationary state within about $10\,M$. Due to the chosen
gauge conditions and numerical resolutions, we lose track of the
individual and the inner common horizon about $5\,M$ after
coalescence.  This shall be explained in greater detail shortly.


\subsubsection{Numerical details}

We solve the Einstein equations via the Einstein Toolkit
\cite{Loffler:2011ay, EinsteinToolkit:web} in their BSSN formulation
\cite{Alcubierre:2000xu, Alcubierre:2002kk, Brown:2008sb} using the
usual $1+\log$ slicing and $\Gamma$-driver shift conditions.

We set up initial conditions via the puncture method
\cite{Ansorg:2004ds}. We locate apparent horizons via the method
described in \cite{Thornburg:1995cp, Thornburg:2003sf}. The algorithms
to evaluate quantities for isolated and dynamical horizons were
previously described in \cite{Dreyer:2002mx} and
\cite{Schnetter:2006yt}.


We use a domain with an outer boundary at $240\,M$, making use of the
reflection symmetry about the $z=0$ plane and the equal mass
$\pi$-rotation symmetry about the $z$ axis for a domain extent of
$[0; 240] \times [-240; 240] \times[0; 240]$. We employ adaptive mesh
refinement (AMR), tracking the individual and outer common horizon,
and placing a stack of progressively refined regions around these. For
completeness, we list our evolution parameters in table \ref{tab:evo}.
\begin{table}
  \begin{tabular}{lll}
    Parameter & Symbol & Value \\
    \hline
    $1+\log$ & $n$ & $1$ \\
    $1+\log$ & $f$ & $2$ \\
    $\Gamma$-driver & $\eta$ & $0.75$ \\
    $\Gamma$-driver & $F$ & $0.75 \cdot \alpha$ \\
    \hline
    domain radius & & $240$ \\
    AMR levels & & $7$ total \\
    indiv. BH AMR level radii & & $[32, 16, 8, 4, 2, 1]$ \\
    common BH AMR level radii & & $[64, 32, 16, 8, 4, 2]$ \\
    finest resolution & $\Delta x$ & $0.015625$ \\
    \hline
    horizon surface resolution & $\Delta\theta$ & $0.02936\,\mathrm{rad}$ \\
    horizon surface resolution & $\Delta\phi$ & $0.02909\,\mathrm{rad}$ \\
  \end{tabular}
  \caption{Evolution parameters}
  \label{tab:evo}
\end{table}

\section{The area increase law}
\label{sec:finalbh}

While this paper is mainly concerned with the properties of the final
horizon and its approach to equilibrium, it is interesting to start
with a somewhat different issue, namely to understand the various
kinds of horizons present in a binary black hole system and how their
areas evolve. In particular we shall track the areas, coordinate
shapes and some other physical properties of the horizon areas
starting from the two initial horizons right up to when the final
horizon reaches equilibrium.  These questions were also studied in
\cite{Schnetter:2006yt}, but with much better numerics we are now able
to evolve through the merger al the way to the equilibrium state at
late times.

\begin{figure}
\includegraphics[width=0.45\textwidth]{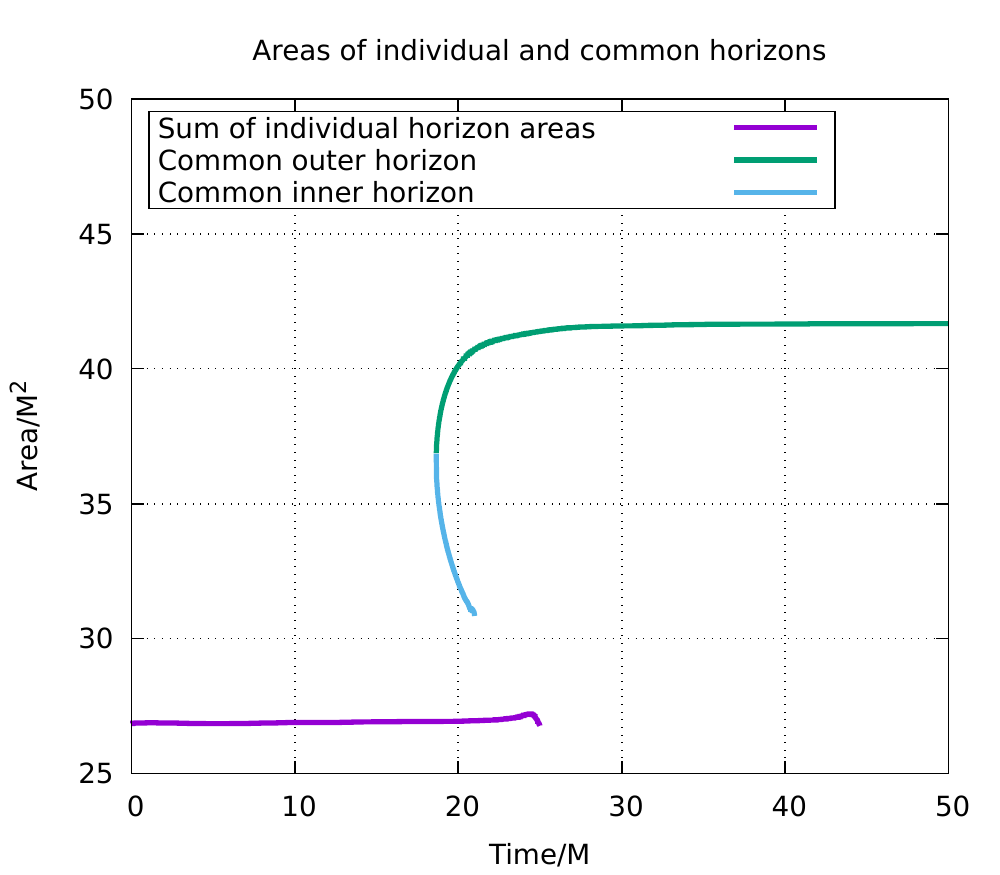}
\caption{This figure shows, as a function of time, the areas of the
  common-outer and inner horizons, and the sum of the areas of the two
  individual horizons.  The common horizon first appears at the
  intersection of the green and blue curves which then splits into an
  inner (blue) and outer (green) horizon.  The area of the inner
  horizon decreases in area while the outer horizon increases and
  eventually reaches an equilibrium value.  The purple curve refers to
  the sum of the areas of the two separate individual horizons.  We
  have plotted only up to $t=50M$ since the area of the common-outer
  horizon after that point is essentially constant. The small dip in
  the area of the individual horizons at the end indicates a problem
  with numerical accuracy for locating the inner horizon, and is not
  to be trusted at that point. }\label{fig:area}
\end{figure}

As mentioned earlier, the common apparent horizon forms at about
$t\approx 18.656M$ in the simulation time.  The common horizon splits
into inner and outer components.  The areas are shown in
Fig.~\ref{fig:area}. The outer horizon continues to grow while the
inner horizon shrinks. The areas of the two individual horizons are
seen to remain essentially constant.  

Fig.~\ref{fig:area} leads us to conjecture that there could be a
3-dimensional marginally trapped tube that interpolates between the
two initial horizons and the final outer marginally trapped tube.  For
this to happen, the curve in Fig.~\ref{fig:area} for the individual
horizons must join with the curve for the inner common horizon whose
area is rapidly decreasing.  If this were to happen, we could obtain
the area as a monotonic function on the smooth three dimensional
surface: start as usual by tracking the individual horizons going
forward in time.  At the point that merger with the common-inner
horizon happens, then we would continue going backwards in time so
that the area is still increasing.  Finally, as can be seen in
Fig.~\ref{fig:area}, this joins smoothly with the common-outer horizon
which increases in area going forward in time, and eventually reaches
equilibrium.

Our numerical simulations are not able to track the inner common
horizon beyond $t \approx 21M$ because it becomes highly distorted and
is most likely not a star shaped surface at that time in our
simulation \footnote{A star shaped surface has the property that a ray
  from the origin intersects the surface exactly once.  This condition
  depends on the coordinates chosen and is a technical condition
  required for the apparent horizon tracker employed here
  \cite{Thornburg:2003sf}.}.  Other simulations have successfully
followed the evolution of the two individual binary marginally trapped
surfaces for a somewhat longer time \cite{Mosta:2015sga} where it was
shown that the two marginally trapped surfaces can penetrate each
other. Although we are able to follow the two individual horizons
somewhat longer than the inner common horizon, again because of
technical issues we are not able to follow them to the point where
they penetrate each other.

\begin{figure}
\includegraphics[width=0.45\textwidth]{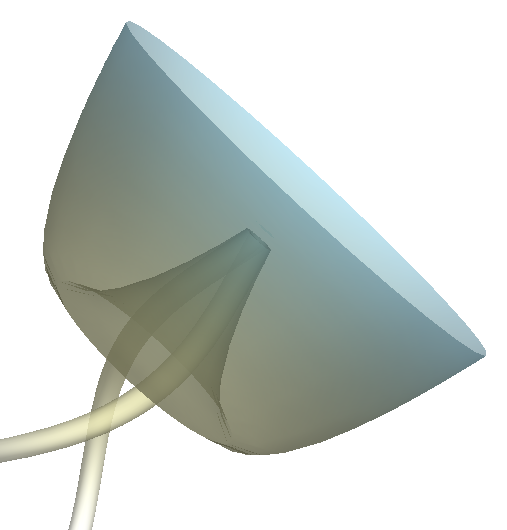}
\caption{A speculative scenario for the fate of the two individual and
  the common-inner horizons. After the common outer and inner horizons
  form, the initial horizons continue to orbit, eventually
  intersecting with one another and then merging with the common inner
  horizon.  Current horizon tracking algorithms are unable to track
  the individual horizons this far into the evolution because of the
  high level of horizon distortion and this scenario remains
  speculation. The plot is not drawn to
  scale.}\label{fig:merger-cartoon}
\end{figure}

The further evolution of both the common inner horizon and the binary
horizons is still unresolved, although it is likely that at some point
they join together. The theorems that guarantee smooth time evolution
of marginally trapped surfaces do not apply in this case because the
general stability conditions do not hold for inner horizons. However,
two main possibilities seem likely. Either, after penetrating one
another the two binary horizons merge together, and then subsequently
merge with the inner horizon. Or, the two horizons merge with the
inner horizon after penetration, but before becoming a single
horizon. An artistic impression (not based on actual data) for the
second possibility is displayed in Fig.~\ref{fig:merger-cartoon}.

If either case were confirmed to be true, one could then introduce a
parameter $\lambda$ along the continuous 3-dimensional surface and the
area $A(\lambda)$ would be a monotonically increasing function
starting from the individual horizons to the final outer horizon which
eventually reaches equilibrium.  It would be of interest to find a
suitable gauge condition which would enable us to track the inner
common-horizon to confirm or disprove this scenario.

The other feature that is obvious from Fig.~\ref{fig:area} is the fact
that the areas of the two individual horizons are essentially constant
even though the start of our simulation is already very close to the
merger.  Only the sum of their areas is shown in Fig.~\ref{fig:area}
but, since we are working with equal mass non-spinning black holes,
the two areas are the same.  There might be interesting effects
related to the tidal interactions between the two black holes, but
that is not the topic for this work.  We instead focus on the final
black hole, i.e. on the inner and outer portions of the common horizon
and its approach to equilibrium.

\begin{figure*}
\includegraphics[width=0.90\textwidth]{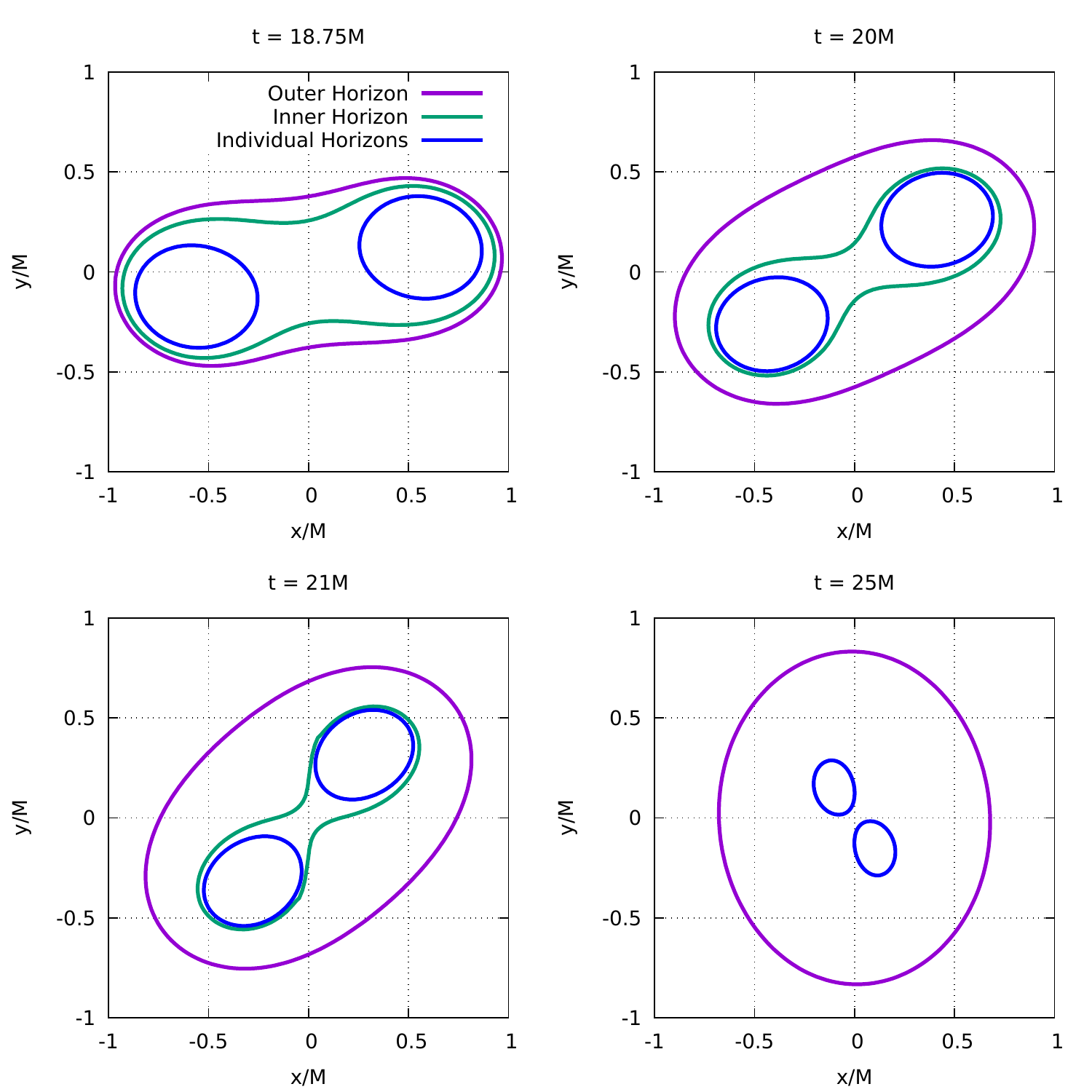}
\caption{The shape of the inner and outer common horizons, and the two
  individual horizons on the equatorial plane at four selected times:
  $t/M=18.75, 20, 21, 25$.  The first ($t=18.75M$) is shortly after
  the common horizon is formed and the third ($t=21M$) is shortly
  before we lose track of the common inner horizon. In the last panel
  ($t=25M$), we are unable to locate the common inner horizon and thus
  only the outer and individual horizons are shown; we lose track of
  the individual horizons soon after this time. In particular, note
  that at $t=21M$, portions of the common inner horizon are almost
  tangential to the y-axis indicating that the inner horizon is close
  to violating the property of being star shaped.  Note also that the
  inner horizons are rapidly decreasing in size in the coordinate
  system used in the simulation which causes the horizon finder to
  lose track of them.  This is a gauge effect and the area of these
  horizons shows no such effect.  Different gauge conditions can be
  used which would make it easier to locate the individual
  horizons. }\label{fig:shape1}
\end{figure*}

\begin{figure}
\includegraphics[width=0.45\textwidth]{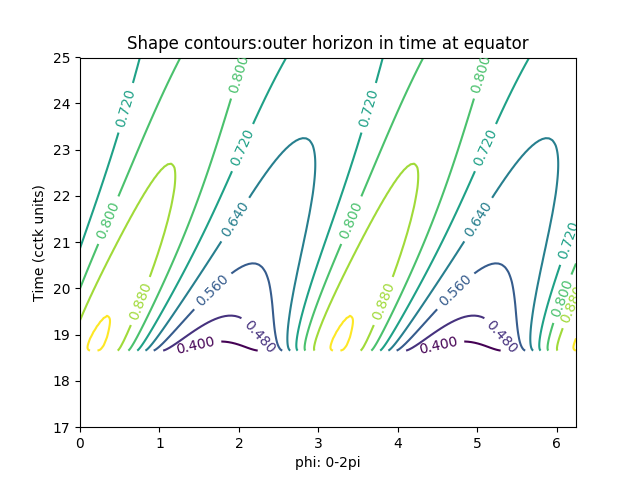}
\caption{The shape of the outer common horizon on the equatorial plane
  as a function of time. See text for explanation. }\label{fig:shape}
\end{figure}

The coordinate shapes of the horizons are depicted in
Figs.~\ref{fig:shape1} and \ref{fig:shape}.  For representing the
horizon in plots, it is convenient to choose sections of the horizon.
Given the presence of reflection symmetry ($z\rightarrow -z$), we
shall use the equatorial plane.  The first set of plots,
Fig.~\ref{fig:shape1}, show the shapes of the horizons on the
equatorial plane at particular times starting just after the formation
of the common horizons, and ending just a short duration before we
lose track of the individual horizons.  We see that the outer horizon
becomes successively more symmetric while the inner horizon becomes
highly asymmetric which makes it difficult for the apparent horizon
tracker to locate it beyond $t\approx 21M$.

Fig.~\ref{fig:shape} shows the outer horizon in more detail. This is a
somewhat unusual way of depicting the evolution but allows us to avoid
showing a large number of two-dimensional plots.  We focus again on
the equatorial plane on which we have polar coordinates $(r,\phi)$ so
that the shape of the horizon can be represented as a radial function
$r(\phi)$. To account for the time evolution we will have a sequence
of functions $r(\phi;t)$. If the horizon were exactly circular, then
$r$ would be constant, but in general it will vary between maximum and
minimum values $r_{max}$ and $r_{min}$ respectively.  We can then
choose a discrete set of values between these extremes and mark, at
each value of $t$, the values of $\phi_i$ where the values $r_i$ are
attained.  Continuing this at different values of $t$, we obtain the
contour plot in the $(\phi,t)$ plane shown in Fig.~\ref{fig:shape} for
the outer common horizon.  The fact that at smaller values of $t$, we
have more allowed values of $r$ means that the horizon has more
irregularities which die away at later times.  For example, at
$t\approx 19M$, the values of $r$ range between about $0.4$ and $0.96$
while at $t\approx 24M$ the range is only between 0.72 and 0.80.  This
does give a useful indication of the horizon shape but it is of course
coordinate dependent. We shall soon use more coordinate independent
geometric multipole moments to quantify how the outer horizon loses
its irregularities at later times.

The area increase gives us an overall picture of the growth of the
horizon.  We can get a detailed picture by looking at the angular
distribution of the flux through the common outer horizon.  We shall
leave a full study of the flux defined in Eq.~(\ref{eq:dhflux}) to a
future study and instead just look at the first term in that
definition, namely the square of the shear.  This term dominates as
the horizon gets closer to equilibrium
\cite{Booth:2003ji,Booth:2006bn}.  However, the null normals defined
in Eq.~(\ref{eq:dhnullnormals}) are not suitable for studying the
approach to equilibrium because as the horizon reaches equilibrium and
becomes null, $\ell^a$ diverges and $n_a$ vanishes.  There is a more
suitable set of null normals used in the simulation.  Consider a
particular Cauchy surface $\Sigma$ containing a MOTS $\mathcal{S}$.
Let $R^a$ be the unit spacelike normal to $\mathcal{S}$ on $\Sigma$,
and let $T^a$ be the unit timelike normal to $\Sigma$.  We define the
null normals 
\begin{equation}
  \bar{\ell}^a = \frac{1}{\sqrt{2}}(T^a+R^a)\,,\quad   \bar{n}^a = \frac{1}{\sqrt{2}}(T^a-R^a)\,. 
\end{equation}
These null normals remain finite throughout the evolution.  There must
then be a function $b$ such that
\begin{equation}
  \bar{\ell}^a = b\ell^a\,,\quad \bar{n}^a = b^{-1}n^a\,,
\end{equation}
and $b\rightarrow 0$ as the horizon approaches equilibrium. The shear
of $\bar{\ell}^a$ scales with $b$: $\bar{\sigma}_{ab} = b\sigma_{ab}$.

\begin{figure}
\includegraphics[width=0.45\textwidth]{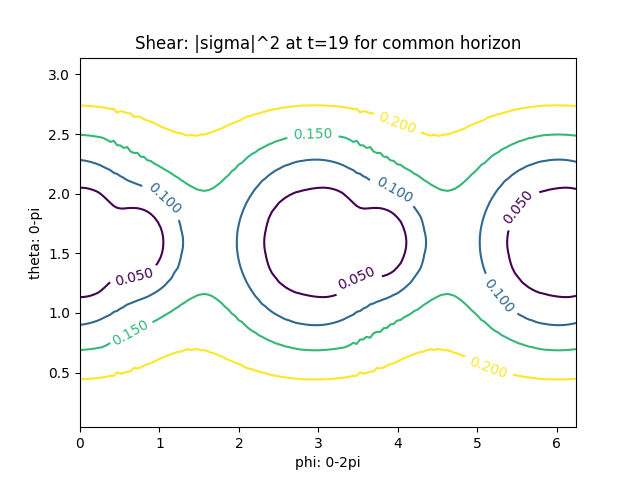}
\includegraphics[width=0.45\textwidth]{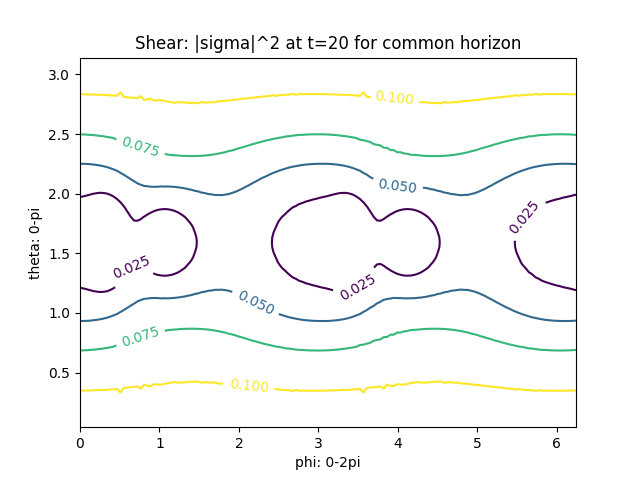}
\includegraphics[width=0.45\textwidth]{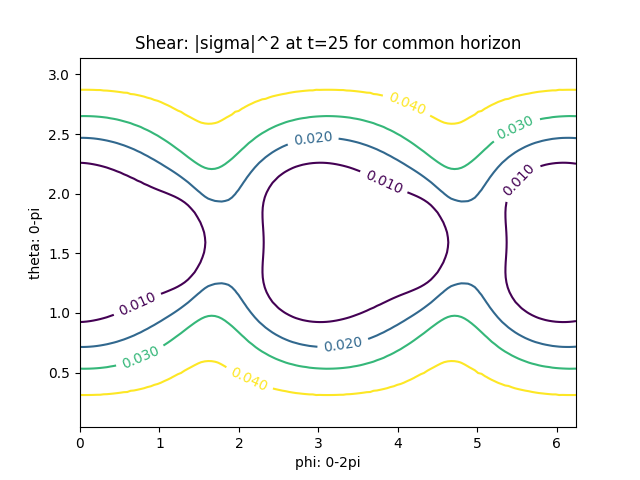}
\caption{The shear at the horizon at $t=19,20,25$.}\label{fig:shear2}
\end{figure}

The modulus of the shear
$|\bar{\sigma}|^2 := \bar{\sigma}_{ab}\bar{\sigma}^{ab}$ on the
horizon at three times, $t/M=19,20,25$, is shown in
Fig.~\ref{fig:shear2} as a function of $(\theta,\phi)$.  As expected,
$|\bar{\sigma}|^2$ decreases with time and moreover, at each time, the
flux is largest through the poles at $\theta =0,\pi$.  We also see
that the horizon shape as shown in Fig.~\ref{fig:shape} has an
apparent rotation (see, for example the slope of, say, the $r=0.720$
contour). This is also clear in the apparent rotation of the horizons
between the panels of Fig.~\ref{fig:shape1}.  On the other hand, the
contour plots of $|\bar{\sigma}|^2$ in Fig.~\ref{fig:shear2} show no
such rotation.  We will discuss further properties of the flux below.

%
%

Other interesting quantities to look at are the expansion of the
in-going null normal, $\Theta_{(n)}$ to the outer common horizon, and
the signatures of the various horizons.  The outer common horizon is,
as expected, spacelike.  The inner horizon becomes partially timelike
soon after it is formed and later completely timelike.  The individual
horizons are null as far as we can tell numerically.  Turning now to
the in-going expansion, recall from Sec.~\ref{subsec:dh} that
$\Theta_{(n)}<0$ is an ingredient of the definition of a dynamical
horizon, and it is used to show that the area of decreases outwards.
As expected, the average of $\Theta_{(n)}$ over a MOTS $\mathcal{S}$,
defined as
\begin{equation}
  \left<\Theta_{(n)}\right>_{\mathcal{S}} := \frac{1}{A_{\mathcal{S}}}\int_{\mathcal{S}} \Theta_{(n)} \widetilde{\epsilon} \,,
\end{equation}
is always negative.  However, this is \emph{not} true point-wise.  In
fact, it turns out that $\Theta_{(n)}$ is completely negative only at
times later than about $\sim 31M$.  Thus, strictly speaking, the world
tube of MOTSs before this time is not a dynamical horizon.  The
essential ingredients of the formalism such as the flux laws,
multipole moments etc. remain valid.  We also point out that the
definition of isolated horizons do not involve any condition on
$\Theta_{(n)}$ and there are several well known examples where it is
not negative everywhere (see
e.g. \cite{Fairhurst:2000xh,Geroch:1982bv}).  These solutions model
situations where a black hole is surrounded by rings of matter which
distort the horizon.  If the binary black hole coalescence were to
occur in the presence of such external matter fields, one would expect
portions of the common horizon to have positive $\Theta_{(n)}$ all the
way through the merger right through to the final equilibrium state.

\begin{figure}
\includegraphics[width=0.45\textwidth]{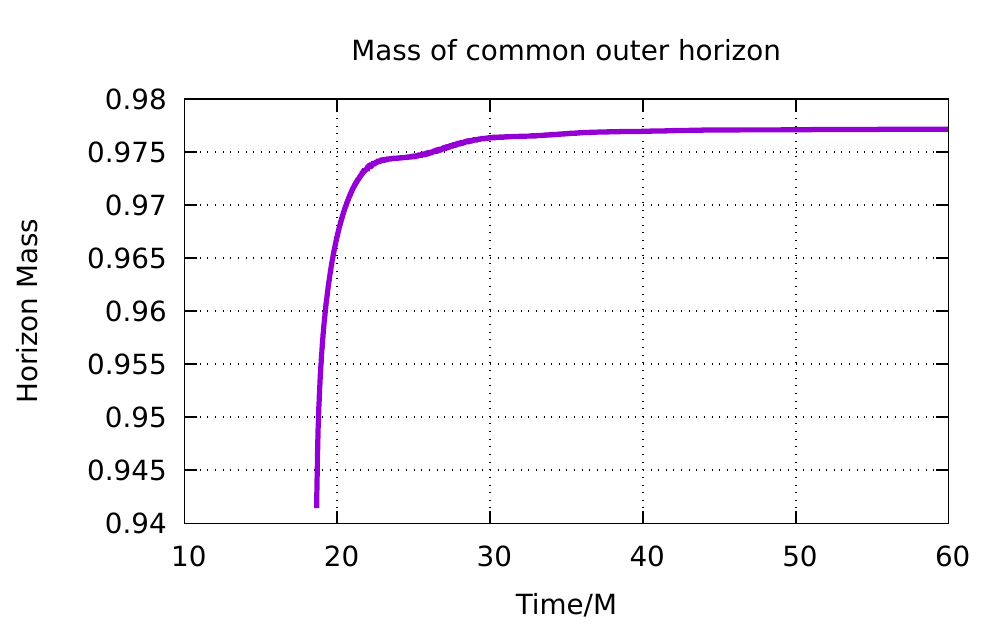}
\includegraphics[width=0.45\textwidth]{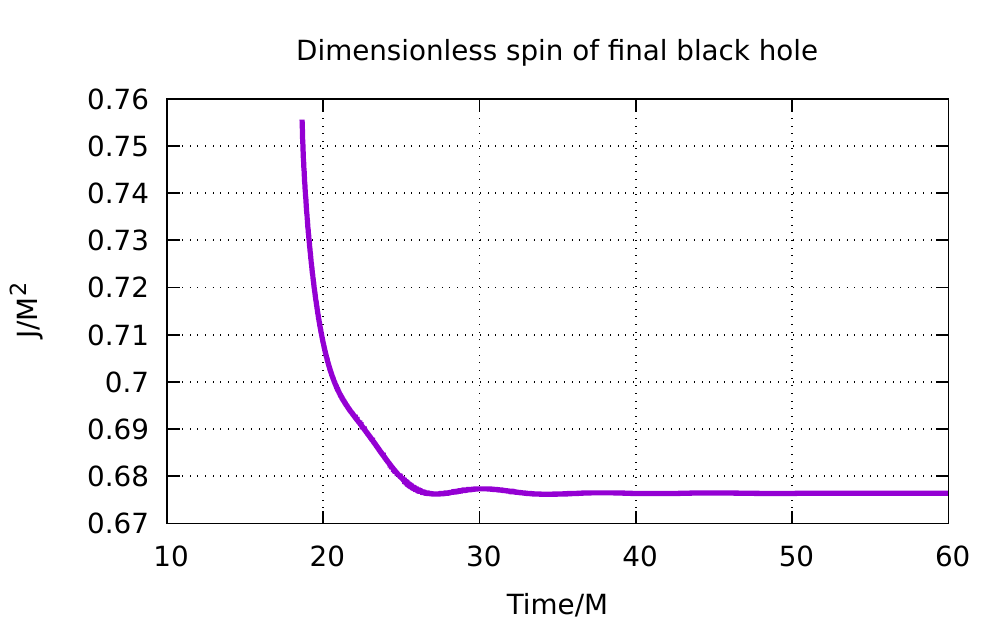}
\caption{The angular momentum and mass of the final black hole.  The
  first figure shows the mass $M(t)$ of the common outer horizon as a
  function of time $t$. The second shows the dimensionless spin,
  i.e. $J(t)/M(t)^2$ with $J$ being the angular
  momentum. }\label{fig:spin_mass}
\end{figure}

Turning now to the physical properties of the horizon, note that the
individual horizons are non-spinning, so their masses are just the
irreducible masses, i.e. $\sqrt{A/16\pi}$ which is completely
determined by the area.  Thus, for angular momentum and mass, only the
common horizon is of interest.  These are shown in
Fig.~\ref{fig:spin_mass} for the common outer horizon. Just like the
area, the mass increases monotonically and reaches an asymptotic value
(there is thus no extraction of energy from the black hole, or
superradiance).  The asymptotic value of the mass is
$M_\infty \sim 0.977$.  The value of the mass at the moment when the
common horizon is formed is $\sim 0.941 M$ and thus the total increase
in the mass of the common- outer horizon is $\Delta M \sim 0.036 M$.
Similarly, the area of the common horizon increases from
$\sim 36.867 M^2$ to $\sim 41.671 M^2$, an increase of
$\Delta A \sim 4.804 M^2$.  We choose to represent the angular
momentum $J$ in terms of the dimensionless quantity
$\chi(t) = J(t)/M^2(t)$.  It can be shown that $\chi$ must always be
less than unity \cite{Jaramillo:2011pg}.  It is seen to decrease with
time, eventually reaching an asymptotic value $\chi \approx 0.68$.
This asymptotic value is consistent with the values found already by
the earliest successful binary black hole simulations
\cite{Pretorius:2005gq,Campanelli:2005dd,Baker:2005vv}.  For later
use, we note that the real and imaginary parts of the angular
frequency of the $n = 0, \ell=m=2$ quasi-normal mode for a Kerr black
hole with this dimensionless spin are
$M\omega^{(0,2,2)} \approx 0.375 -0.089i$.

\section{Approach to equilibrium}
\label{sec:multipoles}

We have already seen how the area, fluxes, mass and spin of the common
horizon evolve and this gives a qualitative picture of the approach to
equilibrium.  To make this more quantitative, we now turn to the mass
and spin multipole moments of the horizon.  This was considered
previously, using somewhat different notions of multipole moments, by
Owen \cite{Owen:2009sb}.  Instead of using an axial vector as done
here, \cite{Owen:2009sb} used eigenfunctions of suitable self-adjoint
operators on the horizons.    

\begin{figure}
\includegraphics[width=0.45\textwidth]{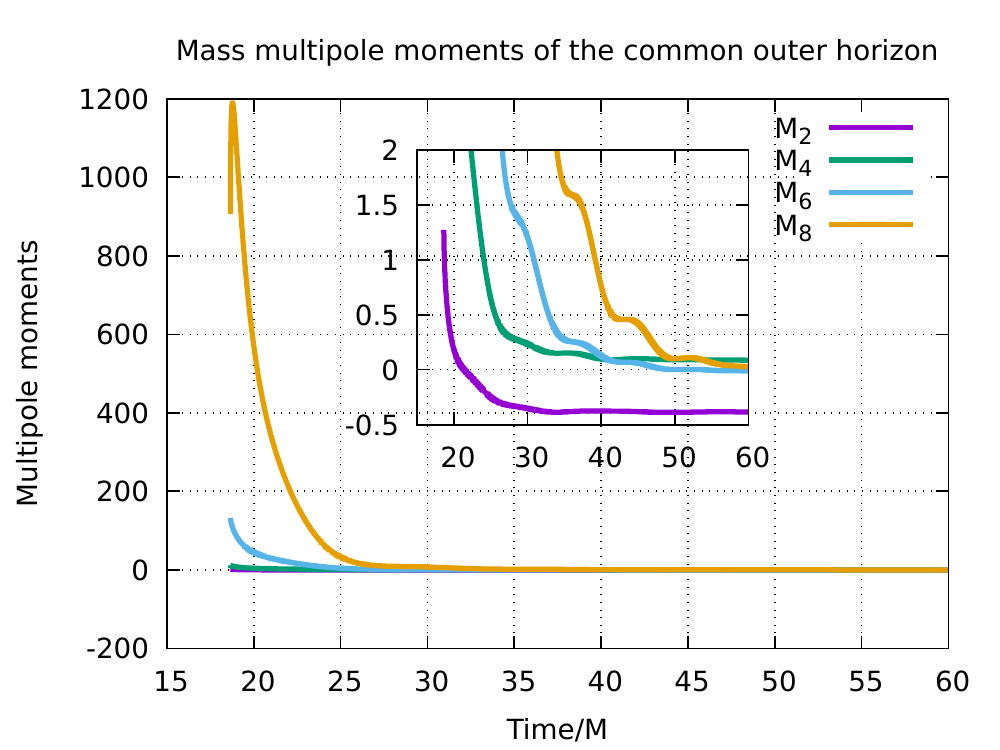}
\includegraphics[width=0.45\textwidth]{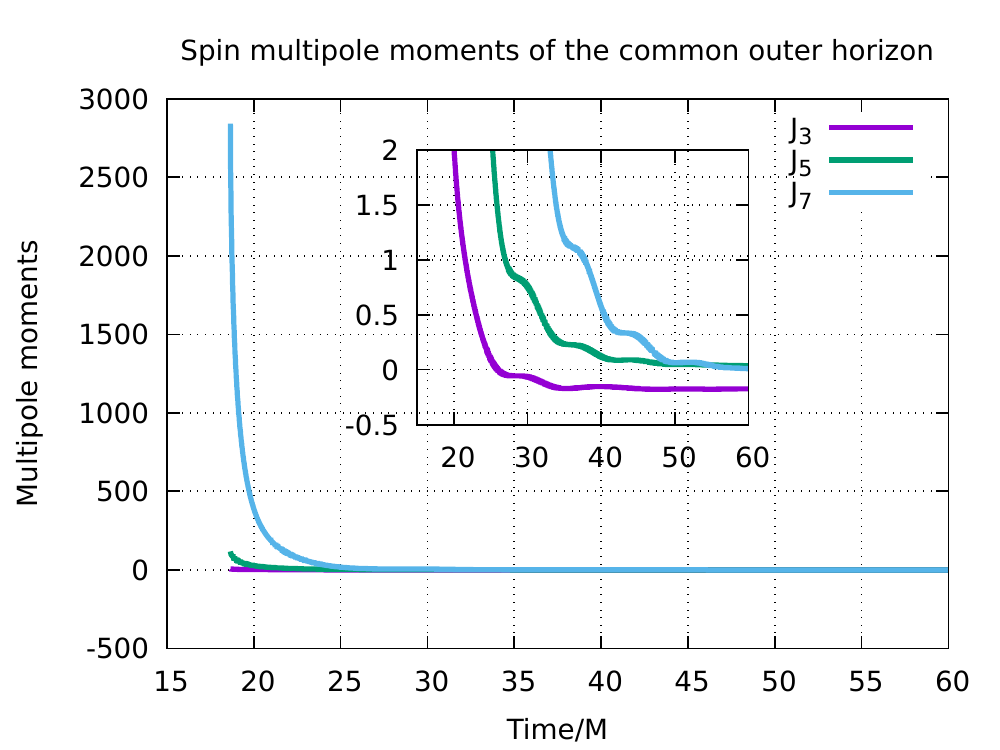}
\caption{The mass and spin moments for the common outer horizon.  The
  inset for both plots is a zoomed version of the bigger plot showing
  better the time variation of the moments.}\label{fig:spin_mass_moments}
\end{figure}

We start by plotting the moments $M_n$ and $J_n$ as functions of time
for the common outer horizon.  Fig.~\ref{fig:spin_mass_moments} shows
the time variation of the mass moments $M_{2,4,6,8}$ and the spin
moments $J_{3,5,7}$. Note that the odd-mass and even-spin moments
vanish due to reflection symmetry.
The first immediate observation about the multipole moments is that
they decay very rapidly to their asymptotic values.  The asymptotic
values of the multipole moments are expected to be the ones of a Kerr
black hole with mass and angular momentum given by $M_0$ and $J_1$
respectively. It is clear from Fig.~\ref{fig:spin_mass_moments} that
for most of the multipole moments there is no difficulty in
identifying the asymptotic value of the multipole moments. The only
exceptions to this are, as shall be clearer on a closer look, $M_8$
and $J_7$ which are harder to compute numerically because the higher
moments require higher angular resolution.

\begin{figure}
\includegraphics[width=0.45\textwidth]{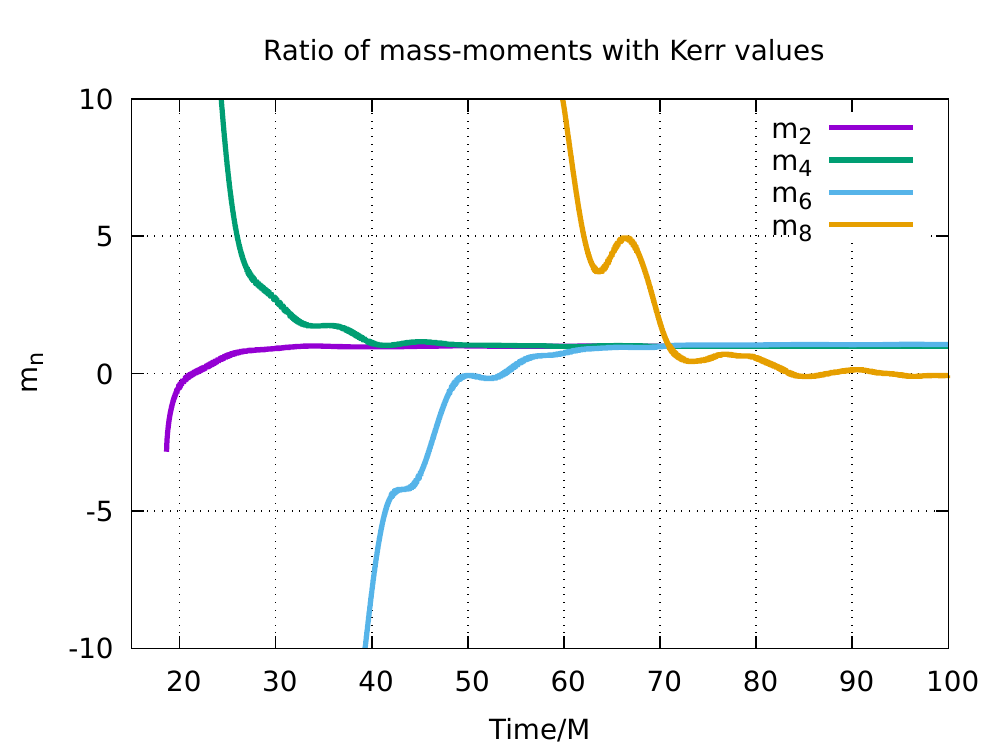}
\includegraphics[width=0.45\textwidth]{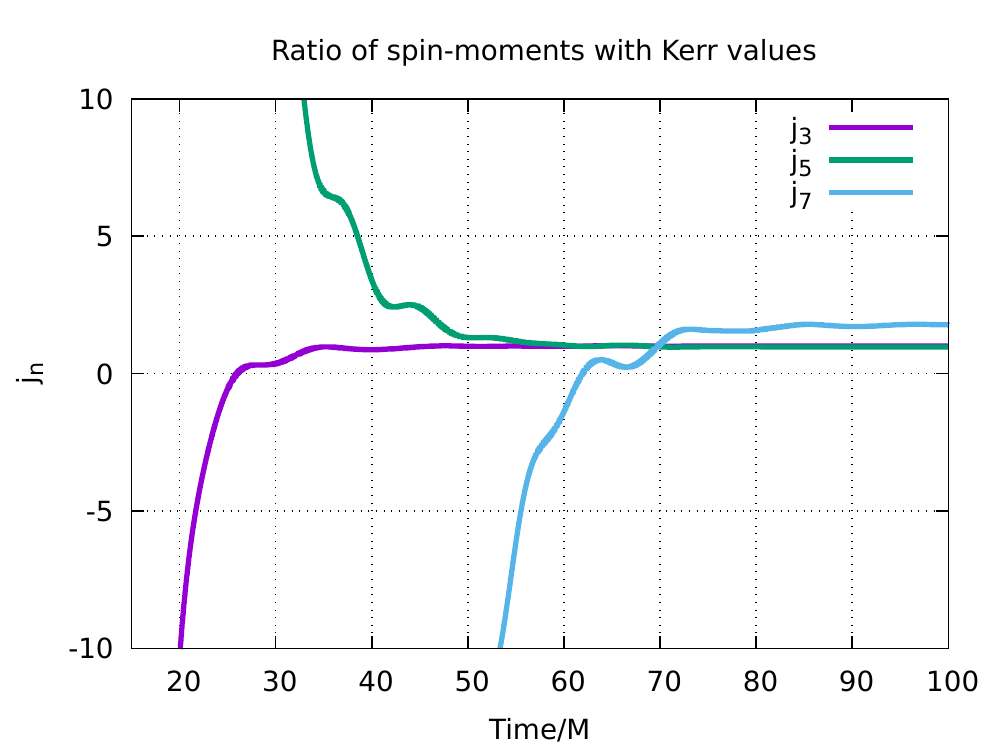}
\caption{The behavior of the ratios of the mass and spin moments to
  the corresponding Kerr moments at each instant of the
  simulation.}\label{fig:moments_kerr_ratios}
\end{figure}

It is instructive to compare the values of the higher multipole
moments at each time to the values a Kerr black hole would have with
the instantaneous values of mass and angular momentum at that time.
It is important to emphasize that these multipoles are different from
the Geroch-Hansen multipole moments defined at spatial infinity.  This
has been considered in \cite{Ashtekar:2004nd} where the differences
between these source multipoles and the field moments are calculated.
For convenience we give in the Appendix expressions for these
multipole moments in terms of the Kerr parameters $M$ and $a$.  At
each time step $t$, given that we have the mass $M_{\mathcal{S}}(t)$
and angular momentum $J_{\mathcal{S}}(t)$, from the expressions in the
appendix, we can calculate $M_n^{\textrm Kerr}(t)$ and
$J_n^{\textrm Kerr}(t)$.  We define the ratios
\begin{equation}
  m_n = \frac{M_n(t)}{M_n^{\textrm Kerr}(t)} \,,\qquad j_n = \frac{J_n(t)}{J_n^{\textrm Kerr}(t)} \,.
\end{equation}
Fig.~\ref{fig:moments_kerr_ratios} shows the behavior of these ratios
with time.  Most moments clearly approach their Kerr values at late
times.  The exceptions to this are $J_7$ and $M_8$ which indicates the
higher numerical errors in calculating the multipole moments beyond
$J_7$ and $M_8$. 

Now we turn to the rate at which the multipole moments decay to their
asymptotic values.  In the linearized theory, the rate at which
perturbations die away is of great physical interest.  This was first
studied by Price \cite{Price:1972pw}.  See
e.g. \cite{Dafermos:2016uzj} for more recent results which proves the
linear stability of Schwarzschild black holes.  The general issue of
the non-linear stability of Kerr black holes is an open question
theoretically speaking.  Numerical simulations offer the possibility
of a better heuristic understanding, and in particular we would like
to investigate whether there are any universalities in the approach to
equilibrium.

It might seem at first glance that the decay is exponential.  Indeed,
one could assume a model of the form
\begin{equation}
\label{eq:expdecay}
  f(t) = f_\infty + Ae^{-\alpha(t-t_0)}\,,
\end{equation}
where $f(t)$ could refer to any of $M_{2,4,6,8}$ or $J_{3,5,7}$, $t_0$
is the time at which the common horizon is formed $t_0 = 18.656M$, and $f_\infty$ is
the asymptotic value of $f$ for large $t$, in this case at $t=100M$.
The parameters $A$ and $\alpha$ could be obtained by fitting the model to the numerically computed data.

\begin{figure}
\centering
\begin{subfigure}{.45\textwidth}
  \centering
  \includegraphics[width=.8\linewidth]{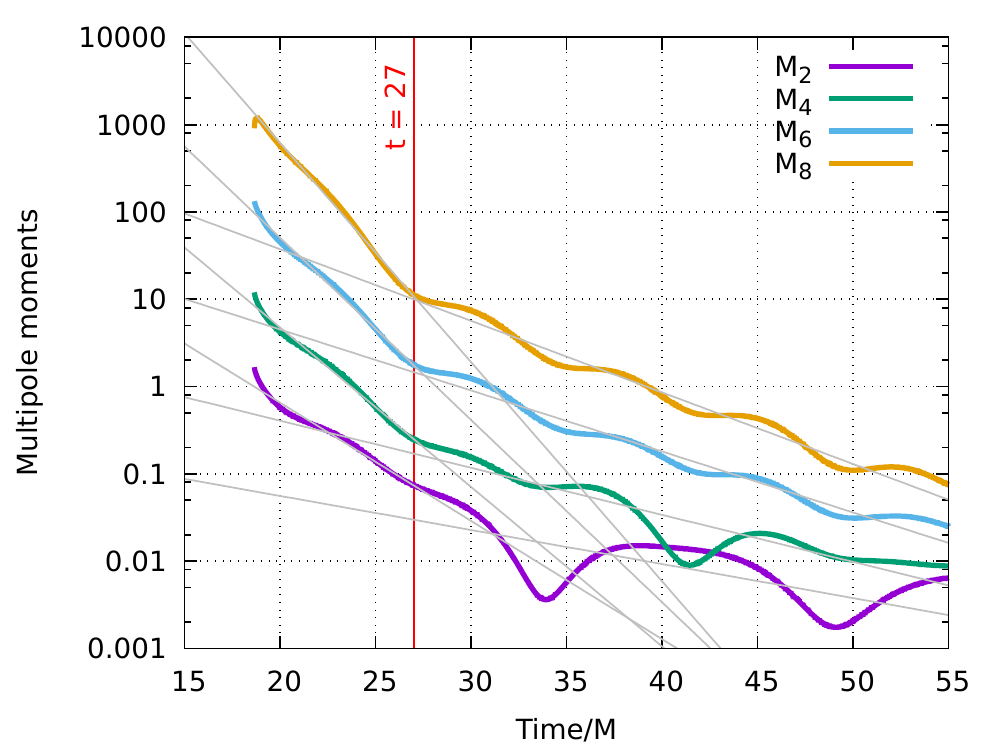}
  \caption{Mass multipoles}
  \label{fig:sub1}
\end{subfigure}\\
\begin{subfigure}{.45\textwidth}
  \centering
  \includegraphics[width=.8\linewidth]{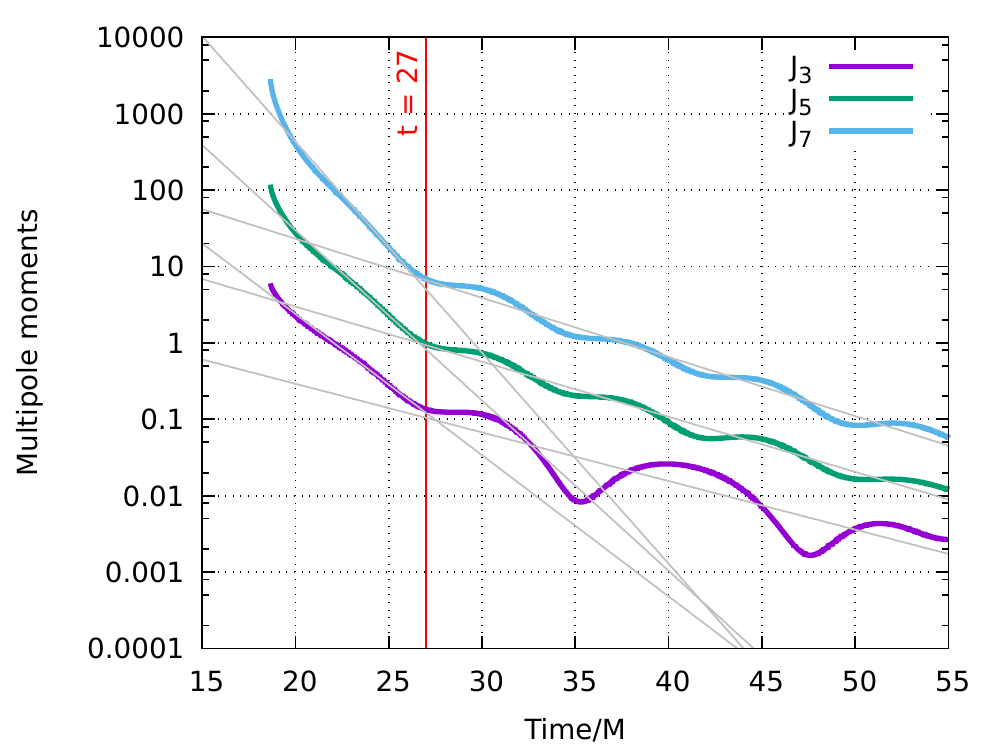}
  \caption{Spin multipoles}
  \label{fig:sub2}
\end{subfigure}
\caption{Logarithmic plot of the mass and spin multipoles. A change in
  the slope is clearly identifiable around the time $27M$. This time
  corresponds to a time approximately $10M$ after the merger and
  delineates a substantial change in the behavior of the horizon. The
  change is most clearly visible in the higher moments but also occurs
  in $M_{2}$ and $J_{3}$.  The figure also shows the best fit straight
  lines to the portion of the plots before and after $t=27M$ (grey
  lines).}
\label{fig:mass_moments_log}
\end{figure}

A closer look reveals that this is in fact not entirely correct.  To
do this, we plot the decay of the multipoles on a logarithmic scale as
shown in Fig.~\ref{fig:mass_moments_log} (the multipole moments have
been appropriately shifted to make them positive at all times, but
still small at late times).  Exponential decay would appear as a
straight line while Fig.~\ref{fig:mass_moments_log} shows different
behavior at early and late times separated at $t\approx 27M$,
approximately $10M$ after the common horizon first forms. Thus, we fit
the multipoles for times $t<27M$ with an exponential decay model using
a simple least squares fitting procedure (we fit the logarithm of the
moments as a linear function of time), and obtain the values of the
decay rates $\alpha$; the results are given in the second column of
Tab.~\ref{tab:fit}.
\begin{table}
  \caption{Best fit values of the exponent $\alpha$ of Eq.~\ref{eq:expdecay} for $M_{2,4,6,8}$ and $J_{3,5,7}$ at early ($t<27M$) and late ($t>27M$) times.}
  \label{tab:fit}
\begin{tabular}{cccc}
 Multipole & $\alpha^{(t<27M)}$ & $\alpha^{(t> 27M)}$\\
  \hline
  $M_2$ &  0.31  & 0.09  \\
  $M_4$ &  0.42  & 0.12  \\
  $M_6$ &  0.48  & 0.16  \\
  $M_8$ &  0.58  & 0.19  \\
  \hline
  $J_3$ &  0.43  & 0.16  \\
  $J_5$ &  0.51  & 0.17  \\
  $J_7$ &  0.64  & 0.18 
\end{tabular}
\end{table}

We now turn to the late time behavior of the multipole moments for
$t>27M$.  Price's law in the linearized context suggests a power-law
fall-off at large times once the exponential part has become
negligible.  Much more likely, in the regime that we are considering,
the moments are linear combinations of exponentially damped functions.
To illustrate the differences from the results of the second column of
Tab.~\ref{tab:fit}, we continue to use the exponential decay model of
Eq.~\ref{eq:expdecay}.  We assume that the moments fall-off as
$e^{-\alpha t}$ within the the range $27<t/M<55$ with the upper value
being chosen arbitrarily (the values do not change significantly when
this is varied). The best fit values of $\alpha$ (again using a
least-squares fit) are shown in the third column of
Tab.~\ref{tab:fit}.

In addition to these decay terms, the oscillation frequencies of the
multipole moments can be determined for $M_{4,6,8}$ and $J_{5,7}$. The
angular frequency can simply be determined by calculating the average
separation between neighboring peaks in $M_n(t)$ and $J_n(t)$ after
the exponential trends have been removed. This yields a value
$M\omega \approx 0.76$ which is roughly twice the dominant
quasi-normal mode frequency.  The steep fall-off of the multipoles
noticeably ends around $t\sim 27M$, roughly $t\approx 10M$ after the
common horizon forms.  This provides additional support, from a very
different viewpoint, with the proposed transition time of
$\approx 10M$ from the merger to the ringdown (after the peak of the
luminosity) found by \cite{Kamaretsos:2011um}.  Caveats to this
conclusion are discussed in Sec.~\ref{sec:conclusions}.


\section{Cross-correlations between the horizon and the waveform}
\label{sec:cross-corr}

In the previous sections we have studied the behavior of the various
horizons which appear in the process of a binary black hole
coalescence.  In particular, we have looked at the growth of the
individual horizons and the approach of the common-outer horizon to a
final Kerr state.  Can any of this information be useful for
understanding observations of gravitational radiation in the wavezone?
Clearly, all of these horizons are hidden behind the event horizon and
thus cannot causally affect any observations outside the event
horizon.  There can however be \emph{correlations} between fields in
the wavezone and the horizon.

\begin{figure*}
\includegraphics[width=0.90\textwidth]{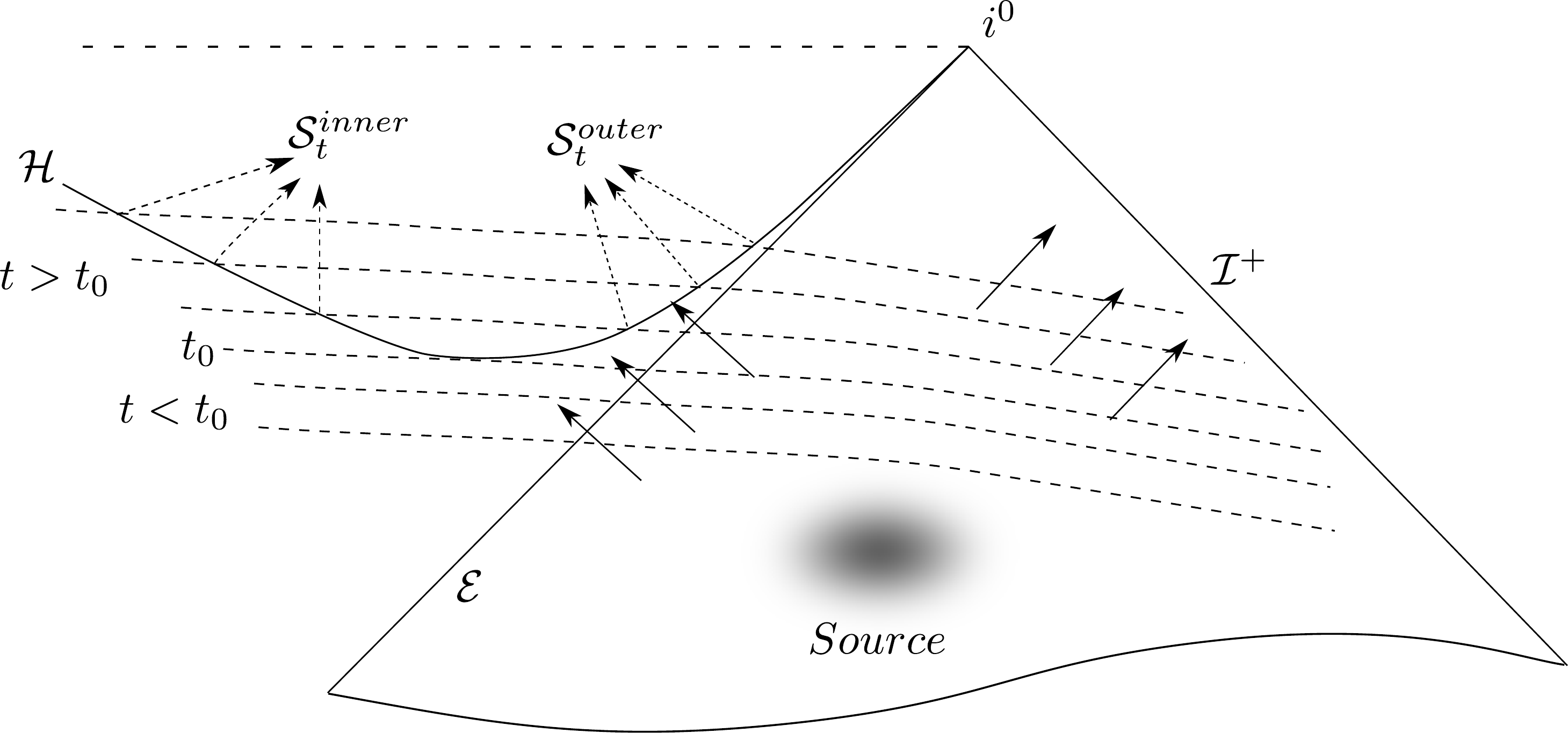}
\caption{A spacetime diagram demonstrating cross-correlations between
  the horizon and null infinity.  Future null infinity is
  $\mathcal{I}^+$ where the gravitational waveform is extracted. The
  event horizon is $\mathcal{E}$, and the dynamical horizon is the
  spacelike surface $\mathcal{H}$. Future timelike infinity is $i^0$
  where $\mathcal{E}$, $\mathcal{H}$, $\mathcal{I}^+$, and the
  singularity (the bold dashed horizontal line) all meet. Cauchy
  surfaces used in the numerical simulation are represented by dashed
  lines.  The common outer horizon is formed at time $t_0$ when the
  Cauchy surface just touches $\mathcal{H}$ in this figure. At later
  times, the intersection of $\mathcal{H}$ with the Cauchy surfaces
  yield the outer and inner marginally trapped surfaces
  $\mathcal{S}_t^{outer}$ and $\mathcal{S}_t^{inner}$
  respectively. The common source is the shaded region which is
  conjectured to lead to correlations between the horizon (either the
  event horizon or preferably the dynamical horizon).  In this
  picture, if the formation of the common horizon is to be correlated
  with the maximum of the outgoing energy flux at $\mathcal{I}^+$, the
  common source for this must be at some earlier time which can
  causally affect both fields at $\mathcal{H}$ and
  $\mathcal{I}^+$. }\label{fig:cc}
\end{figure*}

The intuitive idea of the cross-correlation idea introduced by
Jaramillo et al. is illustrated in Fig.~\ref{fig:cc}.  The figure
shows a portion of spacetime at late times and shows a source which
generically could be due to matter fields or non-linear higher order
contributions due to the gravitational field.  The source will produce
gravitational radiation which can be decomposed into in- and out-going
modes which result respectively in-going flux through the horizon and
outgoing radiation observed at null infinity, or at large distances
from the black hole.  The horizon here could be either the event
horizon, or more conveniently, a dynamical horizon which asymptotes to
the event horizon at future time-like infinity.  It is clear that any
events at the event or dynamical horizon cannot causally affect
observations near null-infinity.  However, given that both are the
result of time evolution of a given initial data set, there could well
be correlations between them.

Analogous to our earlier analysis at the horizon, we now turn our
attention to the wavezone.  Due to their practical importance,
gravitational waveforms have been extensively studied in the
literature.  Regarding the approach of the remnant black hole to
equilibrium, it is found by Kamaretsos et al. \cite{Kamaretsos:2011um}
that the gravitational waveform may be considered to be in the
ringdown phase after a duration $\sim 10M$ following the merger
(defined as the peak of the luminosity); see also
\cite{Thrane:2017lqn} on potential difficulties in ringdown parameter
estimation.  It is interesting that Fig.~\ref{fig:mass_moments_log}
also indicates a time of ~10M after the formation of the common
horizon when the behavior of the horizon multipole moments changes.
Whether this is a mere coincidence or if there is a deeper reason is not
clear at present.  Even if correlations are shown to exist, we have to
deal with the different gauge and coordinate conditions employed at the
horizon and in the wave-zone and it is far from clear how this should
be done. 

\begin{figure}
\includegraphics[width=0.45\textwidth]{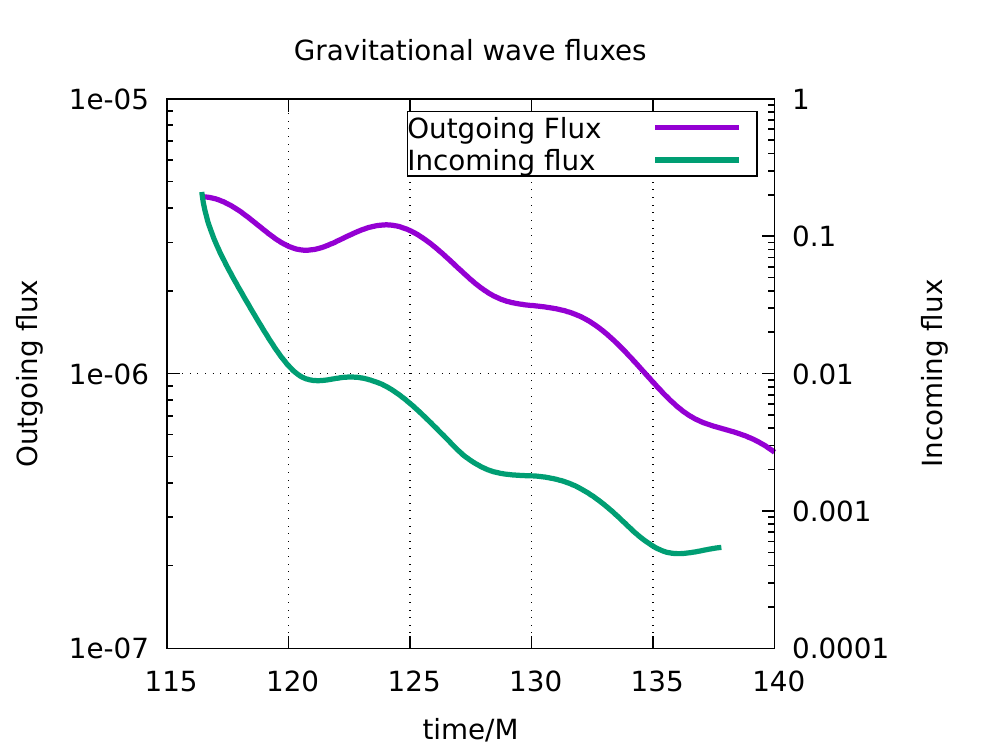}
\caption{The incoming and outgoing fluxes plotted as functions of
`  time. The time axis for the in-going flux $|\bar{\sigma}|^2$ has been
  shifted so that it is aligned with the peak outgoing luminosity.
}\label{fig:fluxes}
\end{figure}

We discuss now additional evidence which lends support to the
existence of such correlations.  As mentioned in the previous
paragraph, \cite{Kamaretsos:2011um} uses the peak luminosity as the
reference time for the merger.  The analog of the luminosity is
precisely the in-going flux through the dynamical horizon discussed
earlier and this is maximum at the moment the common horizon is
formed, consistent with the maximum area growth at the time shown in
Fig.~\ref{fig:area}.  As also suggested in
\cite{Jaramillo:2011rf,Jaramillo:2011re,Jaramillo:2012rr}, we choose
then to compare the shear $|\bar{\sigma}|^2$ at the common horizon
integrated over the horizon, with the luminosity of the $\ell=m=2$
mode of $\Psi_4$. The luminosity of the outgoing radiation is
determined by the News function:
\begin{equation}
  \mathcal{N}^{(\ell, m)}(u) = \int_{-\infty}^u\Psi_4^{(\ell, m)}(u)\,du \,.
\end{equation}
Here we have decomposed the waveform $\Psi_4$ into spin weighted
spherical harmonics and $\Psi_4^{(\ell, m)}$ is the corresponding mode
coefficient as a function of the retarded time $u = t-r$ (appropriate
in the wavezone).  Since we extract the waveform on surfaces at fixed
$r$, we simply take $\Psi_4^{(\ell, m)}$ to be a function of $t$
(starting from the earliest time available in the simulation) and
compare $|\mathcal{N}^{(2,2)}|^2$ with $|\bar{\sigma}|^2$, also as a
function of $t$.  It is worth emphasizing again that even if one
believed in the cross-correlation picture, one would not necessarily
expect a good correlation between the two functions. They are measured
on surfaces at entirely different positions, one inside the event
horizon and one in the wavezone far outside. The gauge condition at
these two surfaces, and thus the meaning of the time coordinate for
the two quantities, do not need to be related with each other in any
way.  Nevertheless, if the change in the behavior of the multipole
moments at $~10M$ after the formation of the common horizon is to be
related to the $~10M$ for the ringdown analysis found by
\cite{Kamaretsos:2011um}, the two must be correlated without adjusting
for any gauge choices.  Let us therefore go ahead and take
$|\mathcal{N}^{(\ell, m)}|^2(t)$ and $|\bar{\sigma}|^2(t)$, shift the time
axis for $|\bar{\sigma}|^2(t)$ so that the two peaks are aligned.  The
result is shown in Fig.~\ref{fig:fluxes}.  By looking at the two
plots, the reader can convince herself that the peaks and troughs of
the two functions are remarkably aligned. This provides further
evidence for the validity of the cross-correlation idea.  The
oscillation frequency of the News function is, as for the horizon
multipoles, twice the frequency of the dominant (i.e.
$n=0, \ell=m=2$) quasi-normal mode.

Finally we consider the angular dependence of $|\bar{\sigma}|^2$ shown
previously in Fig.~\ref{fig:shear2}. These figures show a clear
quadrupolar pattern.  To quantify this we would like to decompose the
shear $\bar{\sigma}$ in terms of spin-weighted spherical harmonics.  In the
Newman-Penrose formalism, we use a null tetrad
$(\ell, n, m, \bar{m})$, with $\ell, n$ being null vectors satisfying
$\ell\cdot n = -1$, $m$ being a complex null vector satisfying
$m\cdot \bar{m} = 1$, and all other inner products vanishing.  Then,
the shear defined in Eq.~\ref{eq:shear-defn} is written as
$\sigma = m^am^b\nabla_a\ell_b$. Under a spin-rotation
$m\rightarrow e^{i\psi}m$, $\sigma$ transforms as
$\sigma\rightarrow e^{2i\psi}\sigma$ and is said to have a
spin-weight 2. Thus, we expect to be able to expand it in terms of
spherical harmonics ${}_2Y^\ell_m(\theta,\phi)$ of spin weight 2
\cite{Goldberg:1966uu,GMS}:
\begin{equation}
  \bar{\sigma} = \sum_{\ell=2}^\infty\sum_{m=-\ell}^\ell \bar{\sigma}^{\ell,m}{}_2Y^\ell_m(\theta,\phi)\,.
\end{equation}
However, just as for the multipole moments, this decomposition
requires a preferred spherical $(\theta,\phi)$ coordinate system and a
suitable area element on the horizon to ensure that the different
${}_sY^\ell_m$ are orthogonal.  Moreover, the proof of being able to
expand tensors in terms of the spin weighted spherical harmonics
relies on the action of the rotation group \cite{GMS} which is not
available for a highly distorted horizon.  Again, as for the multipole
moments, we shall ignore these issues for the moment and simply take
$(\theta,\phi)$ to be the coordinates used on the horizons in the
simulation.  Doing this shows, as expected, that $\bar{\sigma}$ is dominated
by the $\ell=2, m=2$ and $\ell=2, m=-2$ modes with a much smaller
contribution from the $\ell=2, m= 0$ mode, which decreases with time.
As an example, the ratio $|\bar{\sigma}^{2,2}/\bar{\sigma}^{2,0}|$.  At $t=19M$
this has the value $\sim 0.55$, decreasing to $\sim 0.38$ at $t=20M$
and $\sim 0.23$ at $t=25M$.

\section{Conclusions}
\label{sec:conclusions}

The primary goal of this paper is to study the behavior of marginally
trapped surfaces in binary black hole mergers.  The main tools used in
this analysis are the flux formulae and multipole moments.  As seen in
other simulations previously, marginally trapped surfaces are formed
in pairs.  Thus, when the common marginally trapped surface is formed,
we get an outer and inner common marginally trapped surface when the
two black holes get sufficiently close together.  This pair of
marginally trapped surfaces form a smooth quasi-local horizon whose
area increases monotonically outwards.  We have tracked, as far as
possible, the areas of the individual horizons and the common
horizons.  Future work with better gauge conditions and higher
accuracy might succeed in finding the eventual fate of the individual
horizons and the inner horizon.  We have calculated the shear and the
fluxes through the common horizon, leading to a detailed picture of
how the black hole grows and eventually reaches equilibrium.

We have quantitatively studied how the final black hole settles down
to equilibrium.  In particular, we have evaluated the falloff of the
mass and spin multipole moments and we have shown that the final black
hole is Kerr, as expected.  We have quantified the falloff of the
multipole moments and have found that the moments falloff steeply just
after merger, but after about a duration of $10M$ after the merger,
the falloff rate changes to a lower rate. This fact might be useful in
modeling the gravitational wave signal in the merger phase.  These
might provide useful hints for proving the non-linear stability of
Kerr black holes.  We have found that for the QC-0 initial
configuration and using the multipole moments as defined in this paper,
the behavior of the horizon multipole moments under time evolution
changes at an epoch $\sim 10M$ after the formation of the common
horizon.  This is very similar to existing results in the literature
regarding the time at which the post-merger gravitational waveform can
be considered to be in the ringdown phase.  Clearly, this needs to be
explored and understood further and better quantified for a wide rage
of initial configurations.

The fall-off rates of the multipole moments given in
Tab.~\ref{tab:fit} for $t<27M$ are too steep for them to be related to
the $n=0$ quasi-normal modes of the final black hole.  If they are at
all related to quasi-normal ringing, it must be due to the higher
overtones.  See e.g. Fig. 1 of \cite{Dreyer:2003bv} where it is clear
that the imaginary part of the quasi-normal mode frequency is not
greater (in absolute value) than $\sim 0.2$ while the exponents in the
second column of Tab.~\ref{tab:fit} are all greater than $\sim 0.3$.
Alternatively, this might be a genuine non-linear effect unearthed by
using the multipole moments.  However, \cite{Bhagwat:2017tkm} using
the multipole moments defined in \cite{Owen:2009sb}, have found no
such transition in the multipoles.  There could be several reasons for
this. First, note that the multipole moments used here are different
from \cite{Owen:2009sb}.  We have used here the coordinate z-axis (or
equivalently, the axial vector $\varphi^a = \partial_\varphi$) to
define the multipole moments.  This is almost certainly not accurate
just after the merger.  The initial steep fall-off might simply be due
to this choice producing a non-physical effect (the fluxes and the
correlations described in the previous section, which do not depend on
axisymmetry, provide some additional evidence for the choice of $~10M$
for the transition point independent of the choice of $\varphi^a$).
The other reason might be related to the initial configuration that we
have chosen.  It might turn out that both choices of multipole moments
are appropriate, but we have just a fraction of an orbit before
merger.  The additional eccentricity in the initial configuration
might be responsible for exciting higher modes in the initial
post-merger phase. This would require a simulation with a longer
inspiral phase (ideally one tuned to GW150914 or other binary black
hole events) to confirm.  Eventually, these question can be addressed
fully only by a more appropriate choice of multipole moments suited to
fully non-symmetric situations as in \cite{Ashtekar:2013qta}.

Finally we have correlated the behavior of the horizon to the waveform
extracted far away from the black holes.  We have found correlations
between the in-falling and outgoing fluxes both as functions of time
and over angles.  These correlations are unexpected especially in
light of possible differences in the lapse function at the horizon and
at the waveform extraction surface.  This lends additional evidence to
the results of
\cite{Jaramillo:2011rf,Jaramillo:2011re,Jaramillo:2012rr} and might
prove to be a useful tool to observationally study the strong field
region from gravitational wave detections and in gravitational
waveform modeling.  An important aspect of this problem is to find the
free data that can be specified on a dynamical horizon to solve the
Cauchy problem with initial data prescribed on a dynamical horizon.
Thus, in order to reconstruct a relevant portion of spacetime depicted
in Fig.~\ref{fig:cc} we would specify data on (portions of)
$\mathcal{I}^+$ and $\mathcal{H}$.  This would be equivalent to
specifying data on an initial Cauchy surface in the standard way.  For
the case when $\mathcal{H}$ is an isolated horizon the problem has
been solved
\cite{Krishnan:2012bt,Lewandowski:1999zs,Lewandowski:2014nta,Scholtz:2017ttf,Gurlebeck:2018smy}.
Furthermore, the free data on a spherically symmetric dynamical
horizon has been determined by Bartnik and Isenberg
\cite{Bartnik:2005qj}.  The problem of finding the free data on a
general dynamical horizon is yet to be solved.

An important limitation of our approach is the choice of the axial
symmetry vector. Given that we are working with a system of equal-mass
non-spinning black holes, it is appropriate to take the axial vector
on the horizon to be just $\partial_\varphi$, i.e. to assume that the
spin of the final black hole is aligned with the orbital angular
momentum.  This will not be a good approximation in more generic
situations where we would not expect the horizon to have any
symmetries when it is formed.  The method presented in
\cite{Ashtekar:2013qta}, based on finding a suitable class of
divergence free vector fields and assuming that the equilibrium state
is axisymmetric, deals with this general situation and provides
evolution equations for the multipole moments.  Forthcoming work will
implement these ideas.

\acknowledgments

We thank Jose Luis Jaramillo, Frank Ohme, Abhay Ashtekar and Andrey
Shoom for valuable discussions.
This research was supported in part by Perimeter Institute for
Theoretical Physics. Research at Perimeter Institute is supported by
the Government of Canada through the Department of Innovation, Science
and Economic Development Canada, and by the Province of Ontario
through the Ministry of Research, Innovation and Science. A. G. is
supported, in part, by the Navajbai Ratan Tata Trust research grant.
The numerical simulations for this paper were performed on the Perseus
cluster at The Inter-University Centre for Astronomy and Astrophysics,
Pune, India (IUCAA).  We thank Milton Ruiz and Ajay Vibhute in helping
with the necessary computational set up on the Perseus cluster.

\appendix

\section{Expressions for the Kerr multipole moments} 
\label{sec:appendix} 

We start with the expression for the Kerr metric with mass $M$ and
specific angular momentum $a$ in in-going Eddington-Finklestein
coordinates $(v,r,\theta,\phi)$:
\begin{widetext}
\begin{equation}
  ds^2 = -\left(1-\frac{2Mr}{\rho^2}\right)dv^2 + 2dv\,dr - 2a\sin^2\theta dr\,d\varphi 
  - \frac{4aMr\sin^2\theta}{\rho^2} dv\,d\varphi + \rho^2 d\theta^2 +  \frac{\Sigma^2\sin^2\theta}{\rho^2}\,d\varphi^2\,,
\end{equation}
where 
\begin{equation}
  \rho^2 = r^2 + a^2\cos^2\theta\,, \qquad
  \Delta = r^2-2Mr+a^2\,,\qquad 
  \Sigma^2 = (r^2+a^2)\rho^2 + 2a^2Mr\sin^2\theta\,.
\end{equation}
\end{widetext}
The horizon is located at $\Delta = 0$ i.e. at $r=r_+$ such that
$r_+^2 + a^2 = 2Mr_+$.  The volume form on a cross-section of the
horizon ($r=r_+$ and constant $v$) is
$\epsilon = (r_+^2 + a^2)\sin\theta d\theta \wedge d\varphi$. Thus,
the area of the horizon is $A= 4\pi (r_+^2 + a^2)$ and the area radius
is $R = \sqrt{r_+^2+ a^2}$. 

The Weyl tensor component $\Psi_2$ can be shown to be
\cite{Chandrasekhar:1985kt}
\begin{equation}
  \Psi_2 = -\frac{M}{(r-ia\cos\theta)^3}\,.
\end{equation}
This is in fact the only non-vanishing component of the Weyl tensor
for the Kerr spacetime.  The multipole moments are integrals of
$\Psi_2$ which we now define.  For mathematically precise proofs we
refer to \cite{Ashtekar:2004nd} while here our aim is to derive
expressions for the Kerr multipole moments. The analog of $\cos\theta$
on a general axisymmetric horizon is given by an invariant coordinate
$\zeta$ defined as
\begin{equation}
  \partial_a\zeta = R^{-2} \varphi^b \epsilon_{ba}\,.
\end{equation}
In addition we need $\zeta = +1$ at the north pole and $-1$ at the
south pole (the poles being the two points where $\varphi^a$
vanishes).  It is easy to check that for Kerr we have in fact
$\zeta = \cos\theta$.  The mass and spin multipoles are then
respectively
\begin{eqnarray}
  M_n &=& -\frac{MR^n}{2\pi} \oint_S P_n(\zeta)\mathrm{Re}(\Psi_2) d^2S\,,\\
  J_n &=& -\frac{R^{n+1}}{4\pi} \oint_S P_n(\zeta)\mathrm{Im}(\Psi_2) d^2S\,.
\end{eqnarray}
For the Kerr horizon, these become, with $x=a/r_+$,  
\begin{eqnarray}
  M_n &=& 8M^5R^{n-4}\textrm{Re}\int_{-1}^1  \frac{P_n(\zeta)}{(1-ix\zeta)^3} d\zeta\,,   \label{eq:kerrmp}\\
  J_n &=& 4M^4R^{n-3}\textrm{Im} \int_{-1}^1\frac{P_n(\zeta)}{(1-ix\zeta)^3} d\zeta\,.   \label{eq:kerrjp}
\end{eqnarray}
Here we have used 
\begin{equation}
  x = \frac{2Ma}{2Mr_+} = \frac{2Ma}{r_+^2+a^2} = \frac{8\pi J}{A}\,.
\end{equation}
Define $f_n(x)$ to be the integral appearing in these expressions.
From the properties of the Legendre polynomials and $\Psi_2$ under
reflections ($\zeta\rightarrow -\zeta$), it follows that $f_n$ is
automatically real for even $n$ and imaginary for odd $n$.  The
explicit expressions for the integrals are:
\begin{widetext}
\begin{eqnarray}
  f_2(x) &=& \frac{3x + 5x^3 - 3(1+x^2)^2\tan^{-1}x}{x^3(1+x^2)^2}\,,\\
  f_4(x) &=& \frac{15(1+x^2)^2(7+x^2)\tan^{-1}(x) -81x^5 - 190x^3 - 105x}{2x^5(1 + x^2)^2}\,,\\
  f_6(x) &=& \frac{919x^7 + 5103x^5 + 7665x^3 + 3465x -  105(1+x^2)^2(33+18x^2 + x^4)\tan^{-1}(x)}{8x^7(1+x^2)^2}\,,\\
  f_8(x) &=& \frac{315(1+x^2)^2 (143 + 143x^2 + 33x^4 + x^6)\tan^{-1}x - 3781x^9 - 38232x^7 - 109494x^5 - 120120x^3 - 45045x)}{16x^9(1+x^2)^2}\,,\\
  f_3(x) &=& \frac{15(1+x^2)^2\tan^{-1}x - 15x - 25x^3 - 8x^5}{x^4(1+x^2)^2}\,,\\
  f_5(x) &=& \frac{32x^7 + 343x^5 + 630x^3 + 315x - 105(1+x^2)^2(3+x^2)\tan^{-1}(x)}{2x^6(1+x^2)^2}\,,\\
  f_7(x) &=& \frac{315(1+x^2)^2(143+110x^2+15x^4)\tan^{-1}x - 1024x^9- 22923x^7 - 86499x^5-109725x^3 - 45045x}{40x^8(1+x^2)^2}\,.
\end{eqnarray}
Inserting these in the expressions (\ref{eq:kerrmp}) and
(\ref{eq:kerrjp}) yields explicit expressions for $M_n$ and $J_n$.

\end{widetext}

\bibliography{dh}{}

\begin{thebibliography}{75}%
\makeatletter
\providecommand \@ifxundefined [1]{%
 \@ifx{#1\undefined}
}%
\providecommand \@ifnum [1]{%
 \ifnum #1\expandafter \@firstoftwo
 \else \expandafter \@secondoftwo
 \fi
}%
\providecommand \@ifx [1]{%
 \ifx #1\expandafter \@firstoftwo
 \else \expandafter \@secondoftwo
 \fi
}%
\providecommand \natexlab [1]{#1}%
\providecommand \enquote  [1]{``#1''}%
\providecommand \bibnamefont  [1]{#1}%
\providecommand \bibfnamefont [1]{#1}%
\providecommand \citenamefont [1]{#1}%
\providecommand \href@noop [0]{\@secondoftwo}%
\providecommand \href [0]{\begingroup \@sanitize@url \@href}%
\providecommand \@href[1]{\@@startlink{#1}\@@href}%
\providecommand \@@href[1]{\endgroup#1\@@endlink}%
\providecommand \@sanitize@url [0]{\catcode `\\12\catcode `\$12\catcode
  `\&12\catcode `\#12\catcode `\^12\catcode `\_12\catcode `\%12\relax}%
\providecommand \@@startlink[1]{}%
\providecommand \@@endlink[0]{}%
\providecommand \url  [0]{\begingroup\@sanitize@url \@url }%
\providecommand \@url [1]{\endgroup\@href {#1}{\urlprefix }}%
\providecommand \urlprefix  [0]{URL }%
\providecommand \Eprint [0]{\href }%
\providecommand \doibase [0]{http://dx.doi.org/}%
\providecommand \selectlanguage [0]{\@gobble}%
\providecommand \bibinfo  [0]{\@secondoftwo}%
\providecommand \bibfield  [0]{\@secondoftwo}%
\providecommand \translation [1]{[#1]}%
\providecommand \BibitemOpen [0]{}%
\providecommand \bibitemStop [0]{}%
\providecommand \bibitemNoStop [0]{.\EOS\space}%
\providecommand \EOS [0]{\spacefactor3000\relax}%
\providecommand \BibitemShut  [1]{\csname bibitem#1\endcsname}%
\let\auto@bib@innerbib\@empty
\bibitem [{\citenamefont {Pretorius}(2005)}]{Pretorius:2005gq}%
  \BibitemOpen
  \bibfield  {author} {\bibinfo {author} {\bibfnamefont {Frans}\ \bibnamefont
  {Pretorius}},\ }\bibfield  {title} {\enquote {\bibinfo {title} {{Evolution of
  Binary Black Hole Spacetimes}},}\ }\href {\doibase
  10.1103/PhysRevLett.95.121101} {\bibfield  {journal} {\bibinfo  {journal}
  {Phys. Rev. Lett.}\ }\textbf {\bibinfo {volume} {95}},\ \bibinfo {pages}
  {121101} (\bibinfo {year} {2005})},\ \Eprint
  {http://arxiv.org/abs/gr-qc/0507014} {arXiv:gr-qc/0507014} \BibitemShut
  {NoStop}%
\bibitem [{\citenamefont {Campanelli}\ \emph {et~al.}(2006)\citenamefont
  {Campanelli}, \citenamefont {Lousto}, \citenamefont {Marronetti},\ and\
  \citenamefont {Zlochower}}]{Campanelli:2005dd}%
  \BibitemOpen
  \bibfield  {author} {\bibinfo {author} {\bibfnamefont {Manuela}\ \bibnamefont
  {Campanelli}}, \bibinfo {author} {\bibfnamefont {C.~O.}\ \bibnamefont
  {Lousto}}, \bibinfo {author} {\bibfnamefont {P.}~\bibnamefont {Marronetti}},
  \ and\ \bibinfo {author} {\bibfnamefont {Y.}~\bibnamefont {Zlochower}},\
  }\bibfield  {title} {\enquote {\bibinfo {title} {{Accurate evolutions of
  orbiting black-hole binaries without excision}},}\ }\href {\doibase
  10.1103/PhysRevLett.96.111101} {\bibfield  {journal} {\bibinfo  {journal}
  {Phys. Rev. Lett.}\ }\textbf {\bibinfo {volume} {96}},\ \bibinfo {pages}
  {111101} (\bibinfo {year} {2006})},\ \Eprint
  {http://arxiv.org/abs/gr-qc/0511048} {arXiv:gr-qc/0511048 [gr-qc]}
  \BibitemShut {NoStop}%
\bibitem [{\citenamefont {Baker}\ \emph {et~al.}(2006)\citenamefont {Baker},
  \citenamefont {Centrella}, \citenamefont {Choi}, \citenamefont {Koppitz},\
  and\ \citenamefont {van Meter}}]{Baker:2005vv}%
  \BibitemOpen
  \bibfield  {author} {\bibinfo {author} {\bibfnamefont {John~G.}\ \bibnamefont
  {Baker}}, \bibinfo {author} {\bibfnamefont {Joan}\ \bibnamefont {Centrella}},
  \bibinfo {author} {\bibfnamefont {Dae-Il}\ \bibnamefont {Choi}}, \bibinfo
  {author} {\bibfnamefont {Michael}\ \bibnamefont {Koppitz}}, \ and\ \bibinfo
  {author} {\bibfnamefont {James}\ \bibnamefont {van Meter}},\ }\bibfield
  {title} {\enquote {\bibinfo {title} {{Gravitational wave extraction from an
  inspiraling configuration of merging black holes}},}\ }\href {\doibase
  10.1103/PhysRevLett.96.111102} {\bibfield  {journal} {\bibinfo  {journal}
  {Phys. Rev. Lett.}\ }\textbf {\bibinfo {volume} {96}},\ \bibinfo {pages}
  {111102} (\bibinfo {year} {2006})},\ \Eprint
  {http://arxiv.org/abs/gr-qc/0511103} {arXiv:gr-qc/0511103 [gr-qc]}
  \BibitemShut {NoStop}%
\bibitem [{\citenamefont {Baumgarte}\ and\ \citenamefont
  {Shapiro}(2010)}]{BaumgarteShapiroBook}%
  \BibitemOpen
  \bibfield  {author} {\bibinfo {author} {\bibfnamefont {T}~\bibnamefont
  {Baumgarte}}\ and\ \bibinfo {author} {\bibfnamefont {T}~\bibnamefont
  {Shapiro}},\ }\href@noop {} {\emph {\bibinfo {title} {{Numerical Relativity:
  Solving Einstein's Equations on the Computer}}}}\ (\bibinfo  {publisher}
  {Cambridge University Press},\ \bibinfo {year} {2010})\BibitemShut {NoStop}%
\bibitem [{\citenamefont {Shibata}(2016)}]{ShibataBook}%
  \BibitemOpen
  \bibfield  {author} {\bibinfo {author} {\bibfnamefont {M}~\bibnamefont
  {Shibata}},\ }\href@noop {} {\emph {\bibinfo {title} {{Numerical Relativity
  (100 Years of Relativity, Vol 1)}}}}\ (\bibinfo  {publisher} {World
  Scientific},\ \bibinfo {year} {2016})\BibitemShut {NoStop}%
\bibitem [{\citenamefont {Alcubierre}(2008)}]{AlcubierreBook}%
  \BibitemOpen
  \bibfield  {author} {\bibinfo {author} {\bibfnamefont {M}~\bibnamefont
  {Alcubierre}},\ }\href@noop {} {\emph {\bibinfo {title} {Introduction to 3+1
  Numerical Relativity}}}\ (\bibinfo  {publisher} {Oxford University Press},\
  \bibinfo {year} {2008})\BibitemShut {NoStop}%
\bibitem [{\citenamefont {Ashtekar}\ \emph {et~al.}(2004)\citenamefont
  {Ashtekar}, \citenamefont {Engle}, \citenamefont {Paw{\l}owski},\ and\
  \citenamefont {Van Den~Broeck}}]{Ashtekar:2004gp}%
  \BibitemOpen
  \bibfield  {author} {\bibinfo {author} {\bibfnamefont {Abhay}\ \bibnamefont
  {Ashtekar}}, \bibinfo {author} {\bibfnamefont {Jonathan}\ \bibnamefont
  {Engle}}, \bibinfo {author} {\bibfnamefont {Tomasz}\ \bibnamefont
  {Paw{\l}owski}}, \ and\ \bibinfo {author} {\bibfnamefont {Chris}\
  \bibnamefont {Van Den~Broeck}},\ }\bibfield  {title} {\enquote {\bibinfo
  {title} {{Multipole moments of isolated horizons}},}\ }\href {\doibase
  10.1088/0264-9381/21/11/003} {\bibfield  {journal} {\bibinfo  {journal}
  {Class. Quant. Grav.}\ }\textbf {\bibinfo {volume} {21}},\ \bibinfo {pages}
  {2549--2570} (\bibinfo {year} {2004})},\ \Eprint
  {http://arxiv.org/abs/gr-qc/0401114} {arXiv:gr-qc/0401114} \BibitemShut
  {NoStop}%
\bibitem [{\citenamefont {Ashtekar}\ \emph {et~al.}(2013)\citenamefont
  {Ashtekar}, \citenamefont {Campiglia},\ and\ \citenamefont
  {Shah}}]{Ashtekar:2013qta}%
  \BibitemOpen
  \bibfield  {author} {\bibinfo {author} {\bibfnamefont {Abhay}\ \bibnamefont
  {Ashtekar}}, \bibinfo {author} {\bibfnamefont {Miguel}\ \bibnamefont
  {Campiglia}}, \ and\ \bibinfo {author} {\bibfnamefont {Samir}\ \bibnamefont
  {Shah}},\ }\bibfield  {title} {\enquote {\bibinfo {title} {{Dynamical Black
  Holes: Approach to the Final State}},}\ }\href {\doibase
  10.1103/PhysRevD.88.064045} {\bibfield  {journal} {\bibinfo  {journal} {Phys.
  Rev.}\ }\textbf {\bibinfo {volume} {D88}},\ \bibinfo {pages} {064045}
  (\bibinfo {year} {2013})},\ \Eprint {http://arxiv.org/abs/1306.5697}
  {arXiv:1306.5697 [gr-qc]} \BibitemShut {NoStop}%
\bibitem [{\citenamefont {Schnetter}\ \emph {et~al.}(2006)\citenamefont
  {Schnetter}, \citenamefont {Krishnan},\ and\ \citenamefont
  {Beyer}}]{Schnetter:2006yt}%
  \BibitemOpen
  \bibfield  {author} {\bibinfo {author} {\bibfnamefont {Erik}\ \bibnamefont
  {Schnetter}}, \bibinfo {author} {\bibfnamefont {Badri}\ \bibnamefont
  {Krishnan}}, \ and\ \bibinfo {author} {\bibfnamefont {Florian}\ \bibnamefont
  {Beyer}},\ }\bibfield  {title} {\enquote {\bibinfo {title} {{Introduction to
  dynamical horizons in numerical relativity}},}\ }\href {\doibase
  10.1103/PhysRevD.74.024028} {\bibfield  {journal} {\bibinfo  {journal} {Phys.
  Rev.}\ }\textbf {\bibinfo {volume} {D74}},\ \bibinfo {pages} {024028}
  (\bibinfo {year} {2006})},\ \Eprint {http://arxiv.org/abs/gr-qc/0604015}
  {arXiv:gr-qc/0604015} \BibitemShut {NoStop}%
\bibitem [{\citenamefont {Ashtekar}\ and\ \citenamefont
  {Krishnan}(2004)}]{Ashtekar:2004cn}%
  \BibitemOpen
  \bibfield  {author} {\bibinfo {author} {\bibfnamefont {Abhay}\ \bibnamefont
  {Ashtekar}}\ and\ \bibinfo {author} {\bibfnamefont {Badri}\ \bibnamefont
  {Krishnan}},\ }\bibfield  {title} {\enquote {\bibinfo {title} {{Isolated and
  dynamical horizons and their applications}},}\ }\href@noop {} {\bibfield
  {journal} {\bibinfo  {journal} {Living Rev. Rel.}\ }\textbf {\bibinfo
  {volume} {7}},\ \bibinfo {pages} {10} (\bibinfo {year} {2004})},\ \Eprint
  {http://arxiv.org/abs/gr-qc/0407042} {arXiv:gr-qc/0407042} \BibitemShut
  {NoStop}%
\bibitem [{\citenamefont {Booth}(2005)}]{Booth:2005qc}%
  \BibitemOpen
  \bibfield  {author} {\bibinfo {author} {\bibfnamefont {Ivan}\ \bibnamefont
  {Booth}},\ }\bibfield  {title} {\enquote {\bibinfo {title} {{Black hole
  boundaries}},}\ }\href {\doibase 10.1139/p05-063} {\bibfield  {journal}
  {\bibinfo  {journal} {Can. J. Phys.}\ }\textbf {\bibinfo {volume} {83}},\
  \bibinfo {pages} {1073--1099} (\bibinfo {year} {2005})},\ \Eprint
  {http://arxiv.org/abs/gr-qc/0508107} {arXiv:gr-qc/0508107} \BibitemShut
  {NoStop}%
\bibitem [{\citenamefont {Dreyer}\ \emph {et~al.}(2004)\citenamefont {Dreyer},
  \citenamefont {Kelly}, \citenamefont {Krishnan}, \citenamefont {Finn},
  \citenamefont {Garrison},\ and\ \citenamefont
  {Lopez-Aleman}}]{Dreyer:2003bv}%
  \BibitemOpen
  \bibfield  {author} {\bibinfo {author} {\bibfnamefont {Olaf}\ \bibnamefont
  {Dreyer}}, \bibinfo {author} {\bibfnamefont {Bernard~J.}\ \bibnamefont
  {Kelly}}, \bibinfo {author} {\bibfnamefont {Badri}\ \bibnamefont {Krishnan}},
  \bibinfo {author} {\bibfnamefont {Lee~Samuel}\ \bibnamefont {Finn}}, \bibinfo
  {author} {\bibfnamefont {David}\ \bibnamefont {Garrison}}, \ and\ \bibinfo
  {author} {\bibfnamefont {Ramon}\ \bibnamefont {Lopez-Aleman}},\ }\bibfield
  {title} {\enquote {\bibinfo {title} {{Black hole spectroscopy: Testing
  general relativity through gravitational wave observations}},}\ }\href
  {\doibase 10.1088/0264-9381/21/4/003} {\bibfield  {journal} {\bibinfo
  {journal} {Class. Quant. Grav.}\ }\textbf {\bibinfo {volume} {21}},\ \bibinfo
  {pages} {787--804} (\bibinfo {year} {2004})},\ \Eprint
  {http://arxiv.org/abs/gr-qc/0309007} {arXiv:gr-qc/0309007 [gr-qc]}
  \BibitemShut {NoStop}%
\bibitem [{\citenamefont {Abbott}\ \emph {et~al.}(2016)\citenamefont {Abbott}
  \emph {et~al.}}]{TheLIGOScientific:2016src}%
  \BibitemOpen
  \bibfield  {author} {\bibinfo {author} {\bibfnamefont {B.~P.}\ \bibnamefont
  {Abbott}} \emph {et~al.} (\bibinfo {collaboration} {Virgo, LIGO
  Scientific}),\ }\bibfield  {title} {\enquote {\bibinfo {title} {{Tests of
  general relativity with GW150914}},}\ }\href {\doibase
  10.1103/PhysRevLett.116.221101} {\bibfield  {journal} {\bibinfo  {journal}
  {Phys. Rev. Lett.}\ }\textbf {\bibinfo {volume} {116}},\ \bibinfo {pages}
  {221101} (\bibinfo {year} {2016})},\ \Eprint
  {http://arxiv.org/abs/1602.03841} {arXiv:1602.03841 [gr-qc]} \BibitemShut
  {NoStop}%
\bibitem [{\citenamefont {Thrane}\ \emph {et~al.}(2017)\citenamefont {Thrane},
  \citenamefont {Lasky},\ and\ \citenamefont {Levin}}]{Thrane:2017lqn}%
  \BibitemOpen
  \bibfield  {author} {\bibinfo {author} {\bibfnamefont {Eric}\ \bibnamefont
  {Thrane}}, \bibinfo {author} {\bibfnamefont {Paul~D.}\ \bibnamefont {Lasky}},
  \ and\ \bibinfo {author} {\bibfnamefont {Yuri}\ \bibnamefont {Levin}},\
  }\bibfield  {title} {\enquote {\bibinfo {title} {{Challenges testing the
  no-hair theorem with gravitational waves}},}\ }\href {\doibase
  10.1103/PhysRevD.96.102004} {\bibfield  {journal} {\bibinfo  {journal} {Phys.
  Rev.}\ }\textbf {\bibinfo {volume} {D96}},\ \bibinfo {pages} {102004}
  (\bibinfo {year} {2017})},\ \Eprint {http://arxiv.org/abs/1706.05152}
  {arXiv:1706.05152 [gr-qc]} \BibitemShut {NoStop}%
\bibitem [{\citenamefont {Bhagwat}\ \emph {et~al.}(2017)\citenamefont
  {Bhagwat}, \citenamefont {Okounkova}, \citenamefont {Ballmer}, \citenamefont
  {Brown}, \citenamefont {Giesler}, \citenamefont {Scheel},\ and\ \citenamefont
  {Teukolsky}}]{Bhagwat:2017tkm}%
  \BibitemOpen
  \bibfield  {author} {\bibinfo {author} {\bibfnamefont {Swetha}\ \bibnamefont
  {Bhagwat}}, \bibinfo {author} {\bibfnamefont {Maria}\ \bibnamefont
  {Okounkova}}, \bibinfo {author} {\bibfnamefont {Stefan~W.}\ \bibnamefont
  {Ballmer}}, \bibinfo {author} {\bibfnamefont {Duncan~A.}\ \bibnamefont
  {Brown}}, \bibinfo {author} {\bibfnamefont {Matthew}\ \bibnamefont
  {Giesler}}, \bibinfo {author} {\bibfnamefont {Mark~A.}\ \bibnamefont
  {Scheel}}, \ and\ \bibinfo {author} {\bibfnamefont {Saul~A.}\ \bibnamefont
  {Teukolsky}},\ }\bibfield  {title} {\enquote {\bibinfo {title} {{On choosing
  the start time of binary black hole ringdown}},}\ }\href@noop {} {\
  (\bibinfo {year} {2017})},\ \Eprint {http://arxiv.org/abs/1711.00926}
  {arXiv:1711.00926 [gr-qc]} \BibitemShut {NoStop}%
\bibitem [{\citenamefont {Cabero}\ \emph {et~al.}(2017)\citenamefont {Cabero},
  \citenamefont {Capano}, \citenamefont {Fischer-Birnholtz}, \citenamefont
  {Krishnan}, \citenamefont {Nielsen},\ and\ \citenamefont
  {Nitz}}]{Cabero:2017avf}%
  \BibitemOpen
  \bibfield  {author} {\bibinfo {author} {\bibfnamefont {Miriam}\ \bibnamefont
  {Cabero}}, \bibinfo {author} {\bibfnamefont {Collin~D.}\ \bibnamefont
  {Capano}}, \bibinfo {author} {\bibfnamefont {Ofek}\ \bibnamefont
  {Fischer-Birnholtz}}, \bibinfo {author} {\bibfnamefont {Badri}\ \bibnamefont
  {Krishnan}}, \bibinfo {author} {\bibfnamefont {Alex~B.}\ \bibnamefont
  {Nielsen}}, \ and\ \bibinfo {author} {\bibfnamefont {Alex~H.}\ \bibnamefont
  {Nitz}},\ }\bibfield  {title} {\enquote {\bibinfo {title} {{Observational
  tests of the black hole area increase law}},}\ }\href@noop {} {\  (\bibinfo
  {year} {2017})},\ \Eprint {http://arxiv.org/abs/1711.09073} {arXiv:1711.09073
  [gr-qc]} \BibitemShut {NoStop}%
\bibitem [{\citenamefont {Jaramillo}\ \emph
  {et~al.}(2012{\natexlab{a}})\citenamefont {Jaramillo}, \citenamefont
  {Macedo}, \citenamefont {M{\"o}sta},\ and\ \citenamefont
  {Rezzolla}}]{Jaramillo:2011rf}%
  \BibitemOpen
  \bibfield  {author} {\bibinfo {author} {\bibfnamefont {Jose~Luis}\
  \bibnamefont {Jaramillo}}, \bibinfo {author} {\bibfnamefont {Rodrigo~P.}\
  \bibnamefont {Macedo}}, \bibinfo {author} {\bibfnamefont {Philipp}\
  \bibnamefont {M{\"o}sta}}, \ and\ \bibinfo {author} {\bibfnamefont {Luciano}\
  \bibnamefont {Rezzolla}},\ }\bibfield  {title} {\enquote {\bibinfo {title}
  {{Black-hole horizons as probes of black-hole dynamics II: geometrical
  insights}},}\ }\href {\doibase 10.1103/PhysRevD.85.084031} {\bibfield
  {journal} {\bibinfo  {journal} {Phys. Rev.}\ }\textbf {\bibinfo {volume}
  {D85}},\ \bibinfo {pages} {084031} (\bibinfo {year} {2012}{\natexlab{a}})},\
  \Eprint {http://arxiv.org/abs/1108.0061} {arXiv:1108.0061 [gr-qc]}
  \BibitemShut {NoStop}%
\bibitem [{\citenamefont {Jaramillo}\ \emph
  {et~al.}(2012{\natexlab{b}})\citenamefont {Jaramillo}, \citenamefont
  {Macedo}, \citenamefont {M{\"o}sta},\ and\ \citenamefont
  {Rezzolla}}]{Jaramillo:2011re}%
  \BibitemOpen
  \bibfield  {author} {\bibinfo {author} {\bibfnamefont {Jose~Luis}\
  \bibnamefont {Jaramillo}}, \bibinfo {author} {\bibfnamefont
  {Rodrigo~Panosso}\ \bibnamefont {Macedo}}, \bibinfo {author} {\bibfnamefont
  {Philipp}\ \bibnamefont {M{\"o}sta}}, \ and\ \bibinfo {author} {\bibfnamefont
  {Luciano}\ \bibnamefont {Rezzolla}},\ }\bibfield  {title} {\enquote {\bibinfo
  {title} {{Black-hole horizons as probes of black-hole dynamics I: post-merger
  recoil in head-on collisions}},}\ }\href {\doibase
  10.1103/PhysRevD.85.084030} {\bibfield  {journal} {\bibinfo  {journal}
  {Phys.Rev.}\ }\textbf {\bibinfo {volume} {D85}},\ \bibinfo {pages} {084030}
  (\bibinfo {year} {2012}{\natexlab{b}})},\ \Eprint
  {http://arxiv.org/abs/1108.0060} {arXiv:1108.0060 [gr-qc]} \BibitemShut
  {NoStop}%
\bibitem [{\citenamefont {Jaramillo}\ \emph
  {et~al.}(2011{\natexlab{a}})\citenamefont {Jaramillo}, \citenamefont
  {Macedo}, \citenamefont {M{\"o}sta},\ and\ \citenamefont
  {Rezzolla}}]{Jaramillo:2012rr}%
  \BibitemOpen
  \bibfield  {author} {\bibinfo {author} {\bibfnamefont {J.~L.}\ \bibnamefont
  {Jaramillo}}, \bibinfo {author} {\bibfnamefont {R.~P.}\ \bibnamefont
  {Macedo}}, \bibinfo {author} {\bibfnamefont {P.}~\bibnamefont {M{\"o}sta}}, \
  and\ \bibinfo {author} {\bibfnamefont {L.}~\bibnamefont {Rezzolla}},\
  }\bibfield  {title} {\enquote {\bibinfo {title} {{Towards a cross-correlation
  approach to strong-field dynamics in Black Hole spacetimes}},}\ }\bibfield
  {booktitle} {\emph {\bibinfo {booktitle} {{Proceedings, Spanish Relativity
  Meeting : Towards new paradigms. (ERE 2011): Madrid, Spain, August
  29-September 2, 2011}}},\ }\href {\doibase 10.1063/1.4734411} {\bibfield
  {journal} {\bibinfo  {journal} {AIP Conf. Proc.}\ }\textbf {\bibinfo {volume}
  {1458}},\ \bibinfo {pages} {158--173} (\bibinfo {year}
  {2011}{\natexlab{a}})},\ \Eprint {http://arxiv.org/abs/1205.3902}
  {arXiv:1205.3902 [gr-qc]} \BibitemShut {NoStop}%
\bibitem [{\citenamefont {Price}\ and\ \citenamefont
  {Thorne}(1986)}]{Price:1986yy}%
  \BibitemOpen
  \bibfield  {author} {\bibinfo {author} {\bibfnamefont {R.~H.}\ \bibnamefont
  {Price}}\ and\ \bibinfo {author} {\bibfnamefont {K.~S.}\ \bibnamefont
  {Thorne}},\ }\bibfield  {title} {\enquote {\bibinfo {title} {{Membrane
  Viewpoint on Black Holes: Properties and Evolution of the Stretched
  Horizon}},}\ }\href {\doibase 10.1103/PhysRevD.33.915} {\bibfield  {journal}
  {\bibinfo  {journal} {Phys. Rev.}\ }\textbf {\bibinfo {volume} {D33}},\
  \bibinfo {pages} {915--941} (\bibinfo {year} {1986})}\BibitemShut {NoStop}%
\bibitem [{\citenamefont {Rezzolla}\ \emph {et~al.}(2010)\citenamefont
  {Rezzolla}, \citenamefont {Macedo},\ and\ \citenamefont
  {Jaramillo}}]{Rezzolla:2010df}%
  \BibitemOpen
  \bibfield  {author} {\bibinfo {author} {\bibfnamefont {Luciano}\ \bibnamefont
  {Rezzolla}}, \bibinfo {author} {\bibfnamefont {Rodrigo~P.}\ \bibnamefont
  {Macedo}}, \ and\ \bibinfo {author} {\bibfnamefont {Jose~Luis}\ \bibnamefont
  {Jaramillo}},\ }\bibfield  {title} {\enquote {\bibinfo {title}
  {{Understanding the 'anti-kick' in the merger of binary black holes}},}\
  }\href {\doibase 10.1103/PhysRevLett.104.221101} {\bibfield  {journal}
  {\bibinfo  {journal} {Phys. Rev. Lett.}\ }\textbf {\bibinfo {volume} {104}},\
  \bibinfo {pages} {221101} (\bibinfo {year} {2010})},\ \Eprint
  {http://arxiv.org/abs/1003.0873} {arXiv:1003.0873 [gr-qc]} \BibitemShut
  {NoStop}%
\bibitem [{\citenamefont {Thornburg}(2004)}]{Thornburg:2003sf}%
  \BibitemOpen
  \bibfield  {author} {\bibinfo {author} {\bibfnamefont {Jonathan}\
  \bibnamefont {Thornburg}},\ }\bibfield  {title} {\enquote {\bibinfo {title}
  {{A Fast Apparent-Horizon Finder for 3-Dimensional Cartesian Grids in
  Numerical Relativity}},}\ }\href {\doibase 10.1088/0264-9381/21/2/026}
  {\bibfield  {journal} {\bibinfo  {journal} {Class. Quant. Grav.}\ }\textbf
  {\bibinfo {volume} {21}},\ \bibinfo {pages} {743--766} (\bibinfo {year}
  {2004})},\ \Eprint {http://arxiv.org/abs/gr-qc/0306056} {arXiv:gr-qc/0306056}
  \BibitemShut {NoStop}%
\bibitem [{\citenamefont {Andersson}\ \emph {et~al.}(2005)\citenamefont
  {Andersson}, \citenamefont {Mars},\ and\ \citenamefont
  {Simon}}]{Andersson:2005gq}%
  \BibitemOpen
  \bibfield  {author} {\bibinfo {author} {\bibfnamefont {Lars}\ \bibnamefont
  {Andersson}}, \bibinfo {author} {\bibfnamefont {Marc}\ \bibnamefont {Mars}},
  \ and\ \bibinfo {author} {\bibfnamefont {Walter}\ \bibnamefont {Simon}},\
  }\bibfield  {title} {\enquote {\bibinfo {title} {{Local existence of
  dynamical and trapping horizons}},}\ }\href {\doibase
  10.1103/PhysRevLett.95.111102} {\bibfield  {journal} {\bibinfo  {journal}
  {Phys.Rev.Lett.}\ }\textbf {\bibinfo {volume} {95}},\ \bibinfo {pages}
  {111102} (\bibinfo {year} {2005})},\ \Eprint
  {http://arxiv.org/abs/gr-qc/0506013} {arXiv:gr-qc/0506013 [gr-qc]}
  \BibitemShut {NoStop}%
\bibitem [{\citenamefont {Andersson}\ \emph {et~al.}(2009)\citenamefont
  {Andersson}, \citenamefont {Mars}, \citenamefont {Metzger},\ and\
  \citenamefont {Simon}}]{Andersson:2008up}%
  \BibitemOpen
  \bibfield  {author} {\bibinfo {author} {\bibfnamefont {Lars}\ \bibnamefont
  {Andersson}}, \bibinfo {author} {\bibfnamefont {Marc}\ \bibnamefont {Mars}},
  \bibinfo {author} {\bibfnamefont {Jan}\ \bibnamefont {Metzger}}, \ and\
  \bibinfo {author} {\bibfnamefont {Walter}\ \bibnamefont {Simon}},\ }\bibfield
   {title} {\enquote {\bibinfo {title} {{The Time evolution of marginally
  trapped surfaces}},}\ }\href {\doibase 10.1088/0264-9381/26/8/085018}
  {\bibfield  {journal} {\bibinfo  {journal} {Class.Quant.Grav.}\ }\textbf
  {\bibinfo {volume} {26}},\ \bibinfo {pages} {085018} (\bibinfo {year}
  {2009})},\ \Eprint {http://arxiv.org/abs/0811.4721} {arXiv:0811.4721 [gr-qc]}
  \BibitemShut {NoStop}%
\bibitem [{\citenamefont {Andersson}\ \emph {et~al.}(2008)\citenamefont
  {Andersson}, \citenamefont {Mars},\ and\ \citenamefont
  {Simon}}]{Andersson:2007fh}%
  \BibitemOpen
  \bibfield  {author} {\bibinfo {author} {\bibfnamefont {Lars}\ \bibnamefont
  {Andersson}}, \bibinfo {author} {\bibfnamefont {Marc}\ \bibnamefont {Mars}},
  \ and\ \bibinfo {author} {\bibfnamefont {Walter}\ \bibnamefont {Simon}},\
  }\bibfield  {title} {\enquote {\bibinfo {title} {{Stability of marginally
  outer trapped surfaces and existence of marginally outer trapped tubes}},}\
  }\href@noop {} {\bibfield  {journal} {\bibinfo  {journal}
  {Adv.Theor.Math.Phys.}\ }\textbf {\bibinfo {volume} {12}} (\bibinfo {year}
  {2008})},\ \Eprint {http://arxiv.org/abs/0704.2889} {arXiv:0704.2889 [gr-qc]}
  \BibitemShut {NoStop}%
\bibitem [{\citenamefont {Booth}\ \emph {et~al.}(2017)\citenamefont {Booth},
  \citenamefont {Kunduri},\ and\ \citenamefont {O'Grady}}]{Booth:2017fob}%
  \BibitemOpen
  \bibfield  {author} {\bibinfo {author} {\bibfnamefont {Ivan}\ \bibnamefont
  {Booth}}, \bibinfo {author} {\bibfnamefont {Hari~K.}\ \bibnamefont
  {Kunduri}}, \ and\ \bibinfo {author} {\bibfnamefont {Anna}\ \bibnamefont
  {O'Grady}},\ }\bibfield  {title} {\enquote {\bibinfo {title} {{Unstable
  marginally outer trapped surfaces in static spherically symmetric
  spacetimes}},}\ }\href {\doibase 10.1103/PhysRevD.96.024059} {\bibfield
  {journal} {\bibinfo  {journal} {Phys. Rev.}\ }\textbf {\bibinfo {volume}
  {D96}},\ \bibinfo {pages} {024059} (\bibinfo {year} {2017})},\ \Eprint
  {http://arxiv.org/abs/1705.03063} {arXiv:1705.03063 [gr-qc]} \BibitemShut
  {NoStop}%
\bibitem [{\citenamefont {Ashtekar}\ and\ \citenamefont
  {Krishnan}(2002)}]{Ashtekar:2002ag}%
  \BibitemOpen
  \bibfield  {author} {\bibinfo {author} {\bibfnamefont {Abhay}\ \bibnamefont
  {Ashtekar}}\ and\ \bibinfo {author} {\bibfnamefont {Badri}\ \bibnamefont
  {Krishnan}},\ }\bibfield  {title} {\enquote {\bibinfo {title} {{Dynamical
  horizons: Energy, angular momentum, fluxes and balance laws}},}\ }\href
  {\doibase 10.1103/PhysRevLett.89.261101} {\bibfield  {journal} {\bibinfo
  {journal} {Phys. Rev. Lett.}\ }\textbf {\bibinfo {volume} {89}},\ \bibinfo
  {pages} {261101} (\bibinfo {year} {2002})},\ \Eprint
  {http://arxiv.org/abs/gr-qc/0207080} {arXiv:gr-qc/0207080} \BibitemShut
  {NoStop}%
\bibitem [{\citenamefont {Ashtekar}\ and\ \citenamefont
  {Krishnan}(2003)}]{Ashtekar:2003hk}%
  \BibitemOpen
  \bibfield  {author} {\bibinfo {author} {\bibfnamefont {Abhay}\ \bibnamefont
  {Ashtekar}}\ and\ \bibinfo {author} {\bibfnamefont {Badri}\ \bibnamefont
  {Krishnan}},\ }\bibfield  {title} {\enquote {\bibinfo {title} {{Dynamical
  horizons and their properties}},}\ }\href {\doibase
  10.1103/PhysRevD.68.104030} {\bibfield  {journal} {\bibinfo  {journal} {Phys.
  Rev.}\ }\textbf {\bibinfo {volume} {D68}},\ \bibinfo {pages} {104030}
  (\bibinfo {year} {2003})},\ \Eprint {http://arxiv.org/abs/gr-qc/0308033}
  {arXiv:gr-qc/0308033} \BibitemShut {NoStop}%
\bibitem [{\citenamefont {Ashtekar}\ \emph
  {et~al.}(2000{\natexlab{a}})\citenamefont {Ashtekar} \emph
  {et~al.}}]{Ashtekar:2000sz}%
  \BibitemOpen
  \bibfield  {author} {\bibinfo {author} {\bibfnamefont {Abhay}\ \bibnamefont
  {Ashtekar}} \emph {et~al.},\ }\bibfield  {title} {\enquote {\bibinfo {title}
  {{Isolated horizons and their applications}},}\ }\href {\doibase
  10.1103/PhysRevLett.85.3564} {\bibfield  {journal} {\bibinfo  {journal}
  {Phys. Rev. Lett.}\ }\textbf {\bibinfo {volume} {85}},\ \bibinfo {pages}
  {3564--3567} (\bibinfo {year} {2000}{\natexlab{a}})},\ \Eprint
  {http://arxiv.org/abs/gr-qc/0006006} {arXiv:gr-qc/0006006} \BibitemShut
  {NoStop}%
\bibitem [{\citenamefont {Ashtekar}\ \emph {et~al.}(1999)\citenamefont
  {Ashtekar}, \citenamefont {Beetle},\ and\ \citenamefont
  {Fairhurst}}]{Ashtekar:1998sp}%
  \BibitemOpen
  \bibfield  {author} {\bibinfo {author} {\bibfnamefont {Abhay}\ \bibnamefont
  {Ashtekar}}, \bibinfo {author} {\bibfnamefont {Christopher}\ \bibnamefont
  {Beetle}}, \ and\ \bibinfo {author} {\bibfnamefont {Stephen}\ \bibnamefont
  {Fairhurst}},\ }\bibfield  {title} {\enquote {\bibinfo {title} {{Isolated
  horizons: A generalization of black hole mechanics}},}\ }\href {\doibase
  10.1088/0264-9381/16/2/027} {\bibfield  {journal} {\bibinfo  {journal}
  {Class. Quant. Grav.}\ }\textbf {\bibinfo {volume} {16}},\ \bibinfo {pages}
  {L1--L7} (\bibinfo {year} {1999})},\ \Eprint
  {http://arxiv.org/abs/gr-qc/9812065} {arXiv:gr-qc/9812065} \BibitemShut
  {NoStop}%
\bibitem [{\citenamefont {Ashtekar}\ \emph
  {et~al.}(2000{\natexlab{b}})\citenamefont {Ashtekar}, \citenamefont
  {Beetle},\ and\ \citenamefont {Fairhurst}}]{Ashtekar:1999yj}%
  \BibitemOpen
  \bibfield  {author} {\bibinfo {author} {\bibfnamefont {Abhay}\ \bibnamefont
  {Ashtekar}}, \bibinfo {author} {\bibfnamefont {Christopher}\ \bibnamefont
  {Beetle}}, \ and\ \bibinfo {author} {\bibfnamefont {Stephen}\ \bibnamefont
  {Fairhurst}},\ }\bibfield  {title} {\enquote {\bibinfo {title} {{Mechanics of
  Isolated Horizons}},}\ }\href {\doibase 10.1088/0264-9381/17/2/301}
  {\bibfield  {journal} {\bibinfo  {journal} {Class. Quant. Grav.}\ }\textbf
  {\bibinfo {volume} {17}},\ \bibinfo {pages} {253--298} (\bibinfo {year}
  {2000}{\natexlab{b}})},\ \Eprint {http://arxiv.org/abs/gr-qc/9907068}
  {arXiv:gr-qc/9907068} \BibitemShut {NoStop}%
\bibitem [{\citenamefont {Ashtekar}\ \emph {et~al.}(2001)\citenamefont
  {Ashtekar}, \citenamefont {Beetle},\ and\ \citenamefont
  {Lewandowski}}]{Ashtekar:2001is}%
  \BibitemOpen
  \bibfield  {author} {\bibinfo {author} {\bibfnamefont {Abhay}\ \bibnamefont
  {Ashtekar}}, \bibinfo {author} {\bibfnamefont {Christopher}\ \bibnamefont
  {Beetle}}, \ and\ \bibinfo {author} {\bibfnamefont {Jerzy}\ \bibnamefont
  {Lewandowski}},\ }\bibfield  {title} {\enquote {\bibinfo {title} {{Mechanics
  of Rotating Isolated Horizons}},}\ }\href {\doibase
  10.1103/PhysRevD.64.044016} {\bibfield  {journal} {\bibinfo  {journal} {Phys.
  Rev.}\ }\textbf {\bibinfo {volume} {D64}},\ \bibinfo {pages} {044016}
  (\bibinfo {year} {2001})},\ \Eprint {http://arxiv.org/abs/gr-qc/0103026}
  {arXiv:gr-qc/0103026} \BibitemShut {NoStop}%
\bibitem [{\citenamefont {Ashtekar}\ \emph {et~al.}(2002)\citenamefont
  {Ashtekar}, \citenamefont {Beetle},\ and\ \citenamefont
  {Lewandowski}}]{Ashtekar:2001jb}%
  \BibitemOpen
  \bibfield  {author} {\bibinfo {author} {\bibfnamefont {Abhay}\ \bibnamefont
  {Ashtekar}}, \bibinfo {author} {\bibfnamefont {Christopher}\ \bibnamefont
  {Beetle}}, \ and\ \bibinfo {author} {\bibfnamefont {Jerzy}\ \bibnamefont
  {Lewandowski}},\ }\bibfield  {title} {\enquote {\bibinfo {title} {{Geometry
  of Generic Isolated Horizons}},}\ }\href {\doibase
  10.1088/0264-9381/19/6/311} {\bibfield  {journal} {\bibinfo  {journal}
  {Class. Quant. Grav.}\ }\textbf {\bibinfo {volume} {19}},\ \bibinfo {pages}
  {1195--1225} (\bibinfo {year} {2002})},\ \Eprint
  {http://arxiv.org/abs/gr-qc/0111067} {arXiv:gr-qc/0111067} \BibitemShut
  {NoStop}%
\bibitem [{\citenamefont {Krishnan}(2014)}]{Krishnan:2013saa}%
  \BibitemOpen
  \bibfield  {author} {\bibinfo {author} {\bibfnamefont {Badri}\ \bibnamefont
  {Krishnan}},\ }\bibfield  {title} {\enquote {\bibinfo {title} {{Quasi-local
  black hole horizons}},}\ }in\ \href {\doibase 10.1007/978-3-642-41992-8_25}
  {\emph {\bibinfo {booktitle} {Springer Handbook of Spacetime}}},\ \bibinfo
  {editor} {edited by\ \bibinfo {editor} {\bibfnamefont {Abhay}\ \bibnamefont
  {Ashtekar}}\ and\ \bibinfo {editor} {\bibfnamefont {Vesselin}\ \bibnamefont
  {Petkov}}}\ (\bibinfo {year} {2014})\ pp.\ \bibinfo {pages} {527--555},\
  \Eprint {http://arxiv.org/abs/1303.4635} {arXiv:1303.4635 [gr-qc]}
  \BibitemShut {NoStop}%
\bibitem [{\citenamefont {Ben-Dov}(2004)}]{BenDov:2004gh}%
  \BibitemOpen
  \bibfield  {author} {\bibinfo {author} {\bibfnamefont {Ishai}\ \bibnamefont
  {Ben-Dov}},\ }\bibfield  {title} {\enquote {\bibinfo {title} {{The Penrose
  inequality and apparent horizons}},}\ }\href {\doibase
  10.1103/PhysRevD.70.124031} {\bibfield  {journal} {\bibinfo  {journal} {Phys.
  Rev.}\ }\textbf {\bibinfo {volume} {D70}},\ \bibinfo {pages} {124031}
  (\bibinfo {year} {2004})},\ \Eprint {http://arxiv.org/abs/gr-qc/0408066}
  {arXiv:gr-qc/0408066 [gr-qc]} \BibitemShut {NoStop}%
\bibitem [{\citenamefont {Booth}\ \emph {et~al.}(2006)\citenamefont {Booth},
  \citenamefont {Brits}, \citenamefont {Gonzalez},\ and\ \citenamefont {Van
  Den~Broeck}}]{Booth:2005ng}%
  \BibitemOpen
  \bibfield  {author} {\bibinfo {author} {\bibfnamefont {Ivan}\ \bibnamefont
  {Booth}}, \bibinfo {author} {\bibfnamefont {Lionel}\ \bibnamefont {Brits}},
  \bibinfo {author} {\bibfnamefont {Jose~A.}\ \bibnamefont {Gonzalez}}, \ and\
  \bibinfo {author} {\bibfnamefont {Chris}\ \bibnamefont {Van Den~Broeck}},\
  }\bibfield  {title} {\enquote {\bibinfo {title} {{Marginally trapped tubes
  and dynamical horizons}},}\ }\href {\doibase 10.1088/0264-9381/23/2/009}
  {\bibfield  {journal} {\bibinfo  {journal} {Class. Quant. Grav.}\ }\textbf
  {\bibinfo {volume} {23}},\ \bibinfo {pages} {413--440} (\bibinfo {year}
  {2006})},\ \Eprint {http://arxiv.org/abs/gr-qc/0506119} {arXiv:gr-qc/0506119
  [gr-qc]} \BibitemShut {NoStop}%
\bibitem [{\citenamefont {Chu}\ \emph {et~al.}(2011)\citenamefont {Chu},
  \citenamefont {Pfeiffer},\ and\ \citenamefont {Cohen}}]{Chu:2010yu}%
  \BibitemOpen
  \bibfield  {author} {\bibinfo {author} {\bibfnamefont {Tony}\ \bibnamefont
  {Chu}}, \bibinfo {author} {\bibfnamefont {Harald~P.}\ \bibnamefont
  {Pfeiffer}}, \ and\ \bibinfo {author} {\bibfnamefont {Michael~I.}\
  \bibnamefont {Cohen}},\ }\bibfield  {title} {\enquote {\bibinfo {title}
  {{Horizon dynamics of distorted rotating black holes}},}\ }\href {\doibase
  10.1103/PhysRevD.83.104018} {\bibfield  {journal} {\bibinfo  {journal} {Phys.
  Rev.}\ }\textbf {\bibinfo {volume} {D83}},\ \bibinfo {pages} {104018}
  (\bibinfo {year} {2011})},\ \Eprint {http://arxiv.org/abs/1011.2601}
  {arXiv:1011.2601 [gr-qc]} \BibitemShut {NoStop}%
\bibitem [{\citenamefont {Hawking}\ and\ \citenamefont
  {Hartle}(1972)}]{Hawking:1972hy}%
  \BibitemOpen
  \bibfield  {author} {\bibinfo {author} {\bibfnamefont {S.~W.}\ \bibnamefont
  {Hawking}}\ and\ \bibinfo {author} {\bibfnamefont {J.~B.}\ \bibnamefont
  {Hartle}},\ }\bibfield  {title} {\enquote {\bibinfo {title} {{Energy and
  angular momentum flow into a black hole}},}\ }\href {\doibase
  10.1007/BF01645515} {\bibfield  {journal} {\bibinfo  {journal} {Commun. Math.
  Phys.}\ }\textbf {\bibinfo {volume} {27}},\ \bibinfo {pages} {283--290}
  (\bibinfo {year} {1972})}\BibitemShut {NoStop}%
\bibitem [{\citenamefont {Booth}\ and\ \citenamefont
  {Fairhurst}(2007)}]{Booth:2006bn}%
  \BibitemOpen
  \bibfield  {author} {\bibinfo {author} {\bibfnamefont {Ivan}\ \bibnamefont
  {Booth}}\ and\ \bibinfo {author} {\bibfnamefont {Stephen}\ \bibnamefont
  {Fairhurst}},\ }\bibfield  {title} {\enquote {\bibinfo {title} {{Isolated,
  slowly evolving, and dynamical trapping horizons: geometry and mechanics from
  surface deformations}},}\ }\href {\doibase 10.1103/PhysRevD.75.084019}
  {\bibfield  {journal} {\bibinfo  {journal} {Phys. Rev.}\ }\textbf {\bibinfo
  {volume} {D75}},\ \bibinfo {pages} {084019} (\bibinfo {year} {2007})},\
  \Eprint {http://arxiv.org/abs/gr-qc/0610032} {arXiv:gr-qc/0610032}
  \BibitemShut {NoStop}%
\bibitem [{\citenamefont {Cabero}\ and\ \citenamefont
  {Krishnan}(2015)}]{Cabero:2014nza}%
  \BibitemOpen
  \bibfield  {author} {\bibinfo {author} {\bibfnamefont {Miriam}\ \bibnamefont
  {Cabero}}\ and\ \bibinfo {author} {\bibfnamefont {Badri}\ \bibnamefont
  {Krishnan}},\ }\bibfield  {title} {\enquote {\bibinfo {title} {{Tidal
  deformations of spinning black holes in Bowen-York initial data}},}\ }\href
  {\doibase 10.1088/0264-9381/32/4/045009} {\bibfield  {journal} {\bibinfo
  {journal} {Class. Quant. Grav.}\ }\textbf {\bibinfo {volume} {32}},\ \bibinfo
  {pages} {045009} (\bibinfo {year} {2015})},\ \Eprint
  {http://arxiv.org/abs/1407.7656} {arXiv:1407.7656 [gr-qc]} \BibitemShut
  {NoStop}%
\bibitem [{\citenamefont {G{\"u}rlebeck}(2015)}]{Gurlebeck:2015xpa}%
  \BibitemOpen
  \bibfield  {author} {\bibinfo {author} {\bibfnamefont {Norman}\ \bibnamefont
  {G{\"u}rlebeck}},\ }\bibfield  {title} {\enquote {\bibinfo {title} {{No-hair
  theorem for Black Holes in Astrophysical Environments}},}\ }\href {\doibase
  10.1103/PhysRevLett.114.151102} {\bibfield  {journal} {\bibinfo  {journal}
  {Phys. Rev. Lett.}\ }\textbf {\bibinfo {volume} {114}},\ \bibinfo {pages}
  {151102} (\bibinfo {year} {2015})},\ \Eprint
  {http://arxiv.org/abs/1503.03240} {arXiv:1503.03240 [gr-qc]} \BibitemShut
  {NoStop}%
\bibitem [{\citenamefont {Dreyer}\ \emph {et~al.}(2003)\citenamefont {Dreyer},
  \citenamefont {Krishnan}, \citenamefont {Shoemaker},\ and\ \citenamefont
  {Schnetter}}]{Dreyer:2002mx}%
  \BibitemOpen
  \bibfield  {author} {\bibinfo {author} {\bibfnamefont {Olaf}\ \bibnamefont
  {Dreyer}}, \bibinfo {author} {\bibfnamefont {Badri}\ \bibnamefont
  {Krishnan}}, \bibinfo {author} {\bibfnamefont {Deirdre}\ \bibnamefont
  {Shoemaker}}, \ and\ \bibinfo {author} {\bibfnamefont {Erik}\ \bibnamefont
  {Schnetter}},\ }\bibfield  {title} {\enquote {\bibinfo {title} {{Introduction
  to Isolated Horizons in Numerical Relativity}},}\ }\href {\doibase
  10.1103/PhysRevD.67.024018} {\bibfield  {journal} {\bibinfo  {journal} {Phys.
  Rev.}\ }\textbf {\bibinfo {volume} {D67}},\ \bibinfo {pages} {024018}
  (\bibinfo {year} {2003})},\ \Eprint {http://arxiv.org/abs/gr-qc/0206008}
  {arXiv:gr-qc/0206008} \BibitemShut {NoStop}%
\bibitem [{\citenamefont {Ashtekar}\ \emph {et~al.}(2005)\citenamefont
  {Ashtekar}, \citenamefont {Engle},\ and\ \citenamefont {Van
  Den~Broeck}}]{Ashtekar:2004nd}%
  \BibitemOpen
  \bibfield  {author} {\bibinfo {author} {\bibfnamefont {Abhay}\ \bibnamefont
  {Ashtekar}}, \bibinfo {author} {\bibfnamefont {Jonathan}\ \bibnamefont
  {Engle}}, \ and\ \bibinfo {author} {\bibfnamefont {Chris}\ \bibnamefont {Van
  Den~Broeck}},\ }\bibfield  {title} {\enquote {\bibinfo {title} {{Quantum
  horizons and black hole entropy: Inclusion of distortion and rotation}},}\
  }\href {\doibase 10.1088/0264-9381/22/4/L02} {\bibfield  {journal} {\bibinfo
  {journal} {Class. Quant. Grav.}\ }\textbf {\bibinfo {volume} {22}},\ \bibinfo
  {pages} {L27--L34} (\bibinfo {year} {2005})},\ \Eprint
  {http://arxiv.org/abs/gr-qc/0412003} {arXiv:gr-qc/0412003} \BibitemShut
  {NoStop}%
\bibitem [{\citenamefont {Cook}\ and\ \citenamefont
  {Whiting}(2007)}]{Cook:2007wr}%
  \BibitemOpen
  \bibfield  {author} {\bibinfo {author} {\bibfnamefont {Gregory~B.}\
  \bibnamefont {Cook}}\ and\ \bibinfo {author} {\bibfnamefont {Bernard~F.}\
  \bibnamefont {Whiting}},\ }\bibfield  {title} {\enquote {\bibinfo {title}
  {{Approximate Killing Vectors on S**2}},}\ }\href {\doibase
  10.1103/PhysRevD.76.041501} {\bibfield  {journal} {\bibinfo  {journal}
  {Phys.Rev.}\ }\textbf {\bibinfo {volume} {D76}},\ \bibinfo {pages} {041501}
  (\bibinfo {year} {2007})},\ \Eprint {http://arxiv.org/abs/0706.0199}
  {arXiv:0706.0199 [gr-qc]} \BibitemShut {NoStop}%
\bibitem [{\citenamefont {Owen}(2007)}]{Owen:2007dya}%
  \BibitemOpen
  \bibfield  {author} {\bibinfo {author} {\bibfnamefont {Robert}\ \bibnamefont
  {Owen}},\ }\emph {\bibinfo {title} {{Topics in numerical relativity : the
  periodic standing-wave approximation, the stability of constraints in free
  evolution, and the spin of dynamical black holes}}},\ \href
  {http://resolver.caltech.edu/CaltechETD:etd-05252007-143511} {Ph.D. thesis},\
  \bibinfo  {school} {Caltech} (\bibinfo {year} {2007})\BibitemShut {NoStop}%
\bibitem [{\citenamefont {Lovelace}\ \emph {et~al.}(2008)\citenamefont
  {Lovelace}, \citenamefont {Owen}, \citenamefont {Pfeiffer},\ and\
  \citenamefont {Chu}}]{Lovelace:2008tw}%
  \BibitemOpen
  \bibfield  {author} {\bibinfo {author} {\bibfnamefont {Geoffrey}\
  \bibnamefont {Lovelace}}, \bibinfo {author} {\bibfnamefont {Robert}\
  \bibnamefont {Owen}}, \bibinfo {author} {\bibfnamefont {Harald~P.}\
  \bibnamefont {Pfeiffer}}, \ and\ \bibinfo {author} {\bibfnamefont {Tony}\
  \bibnamefont {Chu}},\ }\bibfield  {title} {\enquote {\bibinfo {title}
  {{Binary-black-hole initial data with nearly-extremal spins}},}\ }\href
  {\doibase 10.1103/PhysRevD.78.084017} {\bibfield  {journal} {\bibinfo
  {journal} {Phys. Rev.}\ }\textbf {\bibinfo {volume} {D78}},\ \bibinfo {pages}
  {084017} (\bibinfo {year} {2008})},\ \Eprint {http://arxiv.org/abs/0805.4192}
  {arXiv:0805.4192 [gr-qc]} \BibitemShut {NoStop}%
\bibitem [{\citenamefont {Beetle}(2008)}]{Beetle:2008yt}%
  \BibitemOpen
  \bibfield  {author} {\bibinfo {author} {\bibfnamefont {Christopher}\
  \bibnamefont {Beetle}},\ }\href@noop {} {\enquote {\bibinfo {title}
  {{Approximate Killing Fields as an Eigenvalue Problem}},}\ } (\bibinfo {year}
  {2008}),\ \Eprint {http://arxiv.org/abs/0808.1745} {arXiv:0808.1745 [gr-qc]}
  \BibitemShut {NoStop}%
\bibitem [{\citenamefont {Beetle}\ and\ \citenamefont
  {Wilder}(2014)}]{Beetle:2013zga}%
  \BibitemOpen
  \bibfield  {author} {\bibinfo {author} {\bibfnamefont {Christopher}\
  \bibnamefont {Beetle}}\ and\ \bibinfo {author} {\bibfnamefont {Shawn}\
  \bibnamefont {Wilder}},\ }\bibfield  {title} {\enquote {\bibinfo {title}
  {{Perturbative stability of the approximate Killing field eigenvalue
  problem}},}\ }\href {\doibase 10.1088/0264-9381/31/7/075009} {\bibfield
  {journal} {\bibinfo  {journal} {Class. Quant. Grav.}\ }\textbf {\bibinfo
  {volume} {31}},\ \bibinfo {pages} {075009} (\bibinfo {year} {2014})},\
  \Eprint {http://arxiv.org/abs/1401.0074} {arXiv:1401.0074 [gr-qc]}
  \BibitemShut {NoStop}%
\bibitem [{\citenamefont {Baker}\ \emph {et~al.}(2002)\citenamefont {Baker},
  \citenamefont {Campanelli}, \citenamefont {Lousto},\ and\ \citenamefont
  {Takahashi}}]{Baker:2002qf}%
  \BibitemOpen
  \bibfield  {author} {\bibinfo {author} {\bibfnamefont {John~G.}\ \bibnamefont
  {Baker}}, \bibinfo {author} {\bibfnamefont {Manuela}\ \bibnamefont
  {Campanelli}}, \bibinfo {author} {\bibfnamefont {C.~O.}\ \bibnamefont
  {Lousto}}, \ and\ \bibinfo {author} {\bibfnamefont {R.}~\bibnamefont
  {Takahashi}},\ }\bibfield  {title} {\enquote {\bibinfo {title} {{Modeling
  gravitational radiation from coalescing binary black holes}},}\ }\href
  {\doibase 10.1103/PhysRevD.65.124012} {\bibfield  {journal} {\bibinfo
  {journal} {Phys. Rev.}\ }\textbf {\bibinfo {volume} {D65}},\ \bibinfo {pages}
  {124012} (\bibinfo {year} {2002})},\ \Eprint
  {http://arxiv.org/abs/astro-ph/0202469} {arXiv:astro-ph/0202469 [astro-ph]}
  \BibitemShut {NoStop}%
\bibitem [{\citenamefont {Cook}(1994)}]{Cook:1994va}%
  \BibitemOpen
  \bibfield  {author} {\bibinfo {author} {\bibfnamefont {Gregory~B.}\
  \bibnamefont {Cook}},\ }\bibfield  {title} {\enquote {\bibinfo {title}
  {{Three-dimensional initial data for the collision of two black holes II:
  Quasi-circular orbits for equal-mass black holes}},}\ }\href {\doibase
  10.1103/PhysRevD.50.5025} {\bibfield  {journal} {\bibinfo  {journal} {Phys.
  Rev.}\ }\textbf {\bibinfo {volume} {D50}},\ \bibinfo {pages} {5025--5032}
  (\bibinfo {year} {1994})},\ \Eprint {http://arxiv.org/abs/gr-qc/9404043}
  {arXiv:gr-qc/9404043} \BibitemShut {NoStop}%
\bibitem [{\citenamefont {L{\"{o}}ffler}\ \emph {et~al.}(2012)\citenamefont
  {L{\"{o}}ffler}, \citenamefont {Faber}, \citenamefont {Bentivegna},
  \citenamefont {Bode}, \citenamefont {Diener}, \citenamefont {Haas},
  \citenamefont {Hinder}, \citenamefont {Mundim}, \citenamefont {Ott},
  \citenamefont {Schnetter}, \citenamefont {Allen}, \citenamefont
  {Campanelli},\ and\ \citenamefont {Laguna}}]{Loffler:2011ay}%
  \BibitemOpen
  \bibfield  {author} {\bibinfo {author} {\bibfnamefont {Frank}\ \bibnamefont
  {L{\"{o}}ffler}}, \bibinfo {author} {\bibfnamefont {Joshua}\ \bibnamefont
  {Faber}}, \bibinfo {author} {\bibfnamefont {Eloisa}\ \bibnamefont
  {Bentivegna}}, \bibinfo {author} {\bibfnamefont {Tanja}\ \bibnamefont
  {Bode}}, \bibinfo {author} {\bibfnamefont {Peter}\ \bibnamefont {Diener}},
  \bibinfo {author} {\bibfnamefont {Roland}\ \bibnamefont {Haas}}, \bibinfo
  {author} {\bibfnamefont {Ian}\ \bibnamefont {Hinder}}, \bibinfo {author}
  {\bibfnamefont {Bruno~C.}\ \bibnamefont {Mundim}}, \bibinfo {author}
  {\bibfnamefont {Christian~D.}\ \bibnamefont {Ott}}, \bibinfo {author}
  {\bibfnamefont {Erik}\ \bibnamefont {Schnetter}}, \bibinfo {author}
  {\bibfnamefont {Gabrielle}\ \bibnamefont {Allen}}, \bibinfo {author}
  {\bibfnamefont {Manuela}\ \bibnamefont {Campanelli}}, \ and\ \bibinfo
  {author} {\bibfnamefont {Pablo}\ \bibnamefont {Laguna}},\ }\bibfield  {title}
  {\enquote {\bibinfo {title} {{{T}he {E}instein {T}oolkit: {A} {C}ommunity
  {C}omputational {I}nfrastructure for {R}elativistic {A}strophysics}},}\
  }\href {\doibase doi:10.1088/0264-9381/29/11/115001} {\bibfield  {journal}
  {\bibinfo  {journal} {Class. Quantum Grav.}\ }\textbf {\bibinfo {volume}
  {29}},\ \bibinfo {pages} {115001} (\bibinfo {year} {2012})},\ \Eprint
  {http://arxiv.org/abs/arXiv:1111.3344 [gr-qc]} {arXiv:1111.3344 [gr-qc]}
  \BibitemShut {NoStop}%
\bibitem [{EinsteinToolkit()}]{EinsteinToolkit:web}%
  \BibitemOpen
  EinsteinToolkit,\ \href {http://einsteintoolkit.org/} {\enquote {\bibinfo
  {title} {{Einstein Toolkit}: Open software for relativistic astrophysics},}\
  }\BibitemShut {NoStop}%
\bibitem [{\citenamefont {Alcubierre}\ \emph {et~al.}(2000)\citenamefont
  {Alcubierre}, \citenamefont {Allen}, \citenamefont {Br{\"u}gmann},
  \citenamefont {Dramlitsch}, \citenamefont {Font}, \citenamefont
  {Papadopoulos}, \citenamefont {Seidel}, \citenamefont {Stergioulas},
  \citenamefont {Suen},\ and\ \citenamefont {Takahashi}}]{Alcubierre:2000xu}%
  \BibitemOpen
  \bibfield  {author} {\bibinfo {author} {\bibfnamefont {Miguel}\ \bibnamefont
  {Alcubierre}}, \bibinfo {author} {\bibfnamefont {Gabrielle}\ \bibnamefont
  {Allen}}, \bibinfo {author} {\bibfnamefont {Bernd}\ \bibnamefont
  {Br{\"u}gmann}}, \bibinfo {author} {\bibfnamefont {Thomas}\ \bibnamefont
  {Dramlitsch}}, \bibinfo {author} {\bibfnamefont {Jose~A.}\ \bibnamefont
  {Font}}, \bibinfo {author} {\bibfnamefont {Philippos}\ \bibnamefont
  {Papadopoulos}}, \bibinfo {author} {\bibfnamefont {Edward}\ \bibnamefont
  {Seidel}}, \bibinfo {author} {\bibfnamefont {Nikolaos}\ \bibnamefont
  {Stergioulas}}, \bibinfo {author} {\bibfnamefont {Wai-Mo}\ \bibnamefont
  {Suen}}, \ and\ \bibinfo {author} {\bibfnamefont {Ryoji}\ \bibnamefont
  {Takahashi}},\ }\bibfield  {title} {\enquote {\bibinfo {title} {{Towards a
  stable numerical evolution of strongly gravitating systems in general
  relativity: The Conformal treatments}},}\ }\href {\doibase
  10.1103/PhysRevD.62.044034} {\bibfield  {journal} {\bibinfo  {journal} {Phys.
  Rev.}\ }\textbf {\bibinfo {volume} {D62}},\ \bibinfo {pages} {044034}
  (\bibinfo {year} {2000})},\ \Eprint {http://arxiv.org/abs/gr-qc/0003071}
  {arXiv:gr-qc/0003071 [gr-qc]} \BibitemShut {NoStop}%
\bibitem [{\citenamefont {Alcubierre}\ \emph {et~al.}(2003)\citenamefont
  {Alcubierre}, \citenamefont {Br{\"u}gmann}, \citenamefont {Diener},
  \citenamefont {Koppitz}, \citenamefont {Pollney}, \citenamefont {Seidel},\
  and\ \citenamefont {Takahashi}}]{Alcubierre:2002kk}%
  \BibitemOpen
  \bibfield  {author} {\bibinfo {author} {\bibfnamefont {Miguel}\ \bibnamefont
  {Alcubierre}}, \bibinfo {author} {\bibfnamefont {Bernd}\ \bibnamefont
  {Br{\"u}gmann}}, \bibinfo {author} {\bibfnamefont {Peter}\ \bibnamefont
  {Diener}}, \bibinfo {author} {\bibfnamefont {Michael}\ \bibnamefont
  {Koppitz}}, \bibinfo {author} {\bibfnamefont {Denis}\ \bibnamefont
  {Pollney}}, \bibinfo {author} {\bibfnamefont {Edward}\ \bibnamefont
  {Seidel}}, \ and\ \bibinfo {author} {\bibfnamefont {Ryoji}\ \bibnamefont
  {Takahashi}},\ }\bibfield  {title} {\enquote {\bibinfo {title} {{Gauge
  conditions for long term numerical black hole evolutions without
  excision}},}\ }\href {\doibase 10.1103/PhysRevD.67.084023} {\bibfield
  {journal} {\bibinfo  {journal} {Phys. Rev.}\ }\textbf {\bibinfo {volume}
  {D67}},\ \bibinfo {pages} {084023} (\bibinfo {year} {2003})},\ \Eprint
  {http://arxiv.org/abs/gr-qc/0206072} {arXiv:gr-qc/0206072 [gr-qc]}
  \BibitemShut {NoStop}%
\bibitem [{\citenamefont {Brown}\ \emph {et~al.}(2009)\citenamefont {Brown},
  \citenamefont {Diener}, \citenamefont {Sarbach}, \citenamefont {Schnetter},\
  and\ \citenamefont {Tiglio}}]{Brown:2008sb}%
  \BibitemOpen
  \bibfield  {author} {\bibinfo {author} {\bibfnamefont {J.~David}\
  \bibnamefont {Brown}}, \bibinfo {author} {\bibfnamefont {Peter}\ \bibnamefont
  {Diener}}, \bibinfo {author} {\bibfnamefont {Olivier}\ \bibnamefont
  {Sarbach}}, \bibinfo {author} {\bibfnamefont {Erik}\ \bibnamefont
  {Schnetter}}, \ and\ \bibinfo {author} {\bibfnamefont {Manuel}\ \bibnamefont
  {Tiglio}},\ }\bibfield  {title} {\enquote {\bibinfo {title} {{Turduckening
  black holes: an analytical and computational study}},}\ }\href {\doibase
  10.1103/PhysRevD.79.044023} {\bibfield  {journal} {\bibinfo  {journal} {Phys.
  Rev. D}\ }\textbf {\bibinfo {volume} {79}},\ \bibinfo {pages} {044023}
  (\bibinfo {year} {2009})},\ \Eprint {http://arxiv.org/abs/arXiv:0809.3533
  [gr-qc]} {arXiv:0809.3533 [gr-qc]} \BibitemShut {NoStop}%
\bibitem [{\citenamefont {Ansorg}\ \emph {et~al.}(2004)\citenamefont {Ansorg},
  \citenamefont {Br{\"u}gmann},\ and\ \citenamefont {Tichy}}]{Ansorg:2004ds}%
  \BibitemOpen
  \bibfield  {author} {\bibinfo {author} {\bibfnamefont {Marcus}\ \bibnamefont
  {Ansorg}}, \bibinfo {author} {\bibfnamefont {Bernd}\ \bibnamefont
  {Br{\"u}gmann}}, \ and\ \bibinfo {author} {\bibfnamefont {Wolfgang}\
  \bibnamefont {Tichy}},\ }\bibfield  {title} {\enquote {\bibinfo {title} {A
  single-domain spectral method for black hole puncture data},}\ }\href
  {\doibase 10.1103/PhysRevD.70.064011} {\bibfield  {journal} {\bibinfo
  {journal} {Phys. Rev. D}\ }\textbf {\bibinfo {volume} {70}},\ \bibinfo
  {pages} {064011} (\bibinfo {year} {2004})},\ \Eprint
  {http://arxiv.org/abs/arXiv:gr-qc/0404056} {arXiv:gr-qc/0404056} \BibitemShut
  {NoStop}%
\bibitem [{\citenamefont {Thornburg}(1996)}]{Thornburg:1995cp}%
  \BibitemOpen
  \bibfield  {author} {\bibinfo {author} {\bibfnamefont {Jonathan}\
  \bibnamefont {Thornburg}},\ }\bibfield  {title} {\enquote {\bibinfo {title}
  {{Finding apparent horizons in numerical relativity}},}\ }\href {\doibase
  10.1103/PhysRevD.54.4899} {\bibfield  {journal} {\bibinfo  {journal} {Phys.
  Rev. D}\ }\textbf {\bibinfo {volume} {54}},\ \bibinfo {pages} {4899--4918}
  (\bibinfo {year} {1996})},\ \Eprint
  {http://arxiv.org/abs/arXiv:gr-qc/9508014} {arXiv:gr-qc/9508014} \BibitemShut
  {NoStop}%
\bibitem [{\citenamefont {M{\"o}sta}\ \emph {et~al.}(2015)\citenamefont
  {M{\"o}sta}, \citenamefont {Andersson}, \citenamefont {Metzger},
  \citenamefont {Szil{\'a}gyi},\ and\ \citenamefont
  {Winicour}}]{Mosta:2015sga}%
  \BibitemOpen
  \bibfield  {author} {\bibinfo {author} {\bibfnamefont {P.}~\bibnamefont
  {M{\"o}sta}}, \bibinfo {author} {\bibfnamefont {L.}~\bibnamefont
  {Andersson}}, \bibinfo {author} {\bibfnamefont {J.}~\bibnamefont {Metzger}},
  \bibinfo {author} {\bibfnamefont {B.}~\bibnamefont {Szil{\'a}gyi}}, \ and\
  \bibinfo {author} {\bibfnamefont {J.}~\bibnamefont {Winicour}},\ }\bibfield
  {title} {\enquote {\bibinfo {title} {{The Merger of Small and Large Black
  Holes}},}\ }\href {\doibase 10.1088/0264-9381/32/23/235003} {\bibfield
  {journal} {\bibinfo  {journal} {Class. Quant. Grav.}\ }\textbf {\bibinfo
  {volume} {32}},\ \bibinfo {pages} {235003} (\bibinfo {year} {2015})},\
  \Eprint {http://arxiv.org/abs/1501.05358} {arXiv:1501.05358 [gr-qc]}
  \BibitemShut {NoStop}%
\bibitem [{\citenamefont {Booth}\ and\ \citenamefont
  {Fairhurst}(2004)}]{Booth:2003ji}%
  \BibitemOpen
  \bibfield  {author} {\bibinfo {author} {\bibfnamefont {Ivan}\ \bibnamefont
  {Booth}}\ and\ \bibinfo {author} {\bibfnamefont {Stephen}\ \bibnamefont
  {Fairhurst}},\ }\bibfield  {title} {\enquote {\bibinfo {title} {{The first
  law for slowly evolving horizons}},}\ }\href {\doibase
  10.1103/PhysRevLett.92.011102} {\bibfield  {journal} {\bibinfo  {journal}
  {Phys. Rev. Lett.}\ }\textbf {\bibinfo {volume} {92}},\ \bibinfo {pages}
  {011102} (\bibinfo {year} {2004})},\ \Eprint
  {http://arxiv.org/abs/gr-qc/0307087} {arXiv:gr-qc/0307087} \BibitemShut
  {NoStop}%
\bibitem [{\citenamefont {Fairhurst}\ and\ \citenamefont
  {Krishnan}(2001)}]{Fairhurst:2000xh}%
  \BibitemOpen
  \bibfield  {author} {\bibinfo {author} {\bibfnamefont {Stephen}\ \bibnamefont
  {Fairhurst}}\ and\ \bibinfo {author} {\bibfnamefont {Badri}\ \bibnamefont
  {Krishnan}},\ }\bibfield  {title} {\enquote {\bibinfo {title} {{Distorted
  black holes with charge}},}\ }\href {\doibase 10.1142/S0218271801001086}
  {\bibfield  {journal} {\bibinfo  {journal} {Int. J. Mod. Phys.}\ }\textbf
  {\bibinfo {volume} {D10}},\ \bibinfo {pages} {691--710} (\bibinfo {year}
  {2001})},\ \Eprint {http://arxiv.org/abs/gr-qc/0010088} {arXiv:gr-qc/0010088}
  \BibitemShut {NoStop}%
\bibitem [{\citenamefont {Geroch}\ and\ \citenamefont
  {Hartle}(1982)}]{Geroch:1982bv}%
  \BibitemOpen
  \bibfield  {author} {\bibinfo {author} {\bibfnamefont {Robert~P.}\
  \bibnamefont {Geroch}}\ and\ \bibinfo {author} {\bibfnamefont {J.~B.}\
  \bibnamefont {Hartle}},\ }\bibfield  {title} {\enquote {\bibinfo {title}
  {{Distorted black holes}},}\ }\href {\doibase 10.1063/1.525384} {\bibfield
  {journal} {\bibinfo  {journal} {J. Math. Phys.}\ }\textbf {\bibinfo {volume}
  {23}},\ \bibinfo {pages} {680} (\bibinfo {year} {1982})}\BibitemShut
  {NoStop}%
\bibitem [{\citenamefont {Jaramillo}\ \emph
  {et~al.}(2011{\natexlab{b}})\citenamefont {Jaramillo}, \citenamefont
  {Reiris},\ and\ \citenamefont {Dain}}]{Jaramillo:2011pg}%
  \BibitemOpen
  \bibfield  {author} {\bibinfo {author} {\bibfnamefont {Jose~Luis}\
  \bibnamefont {Jaramillo}}, \bibinfo {author} {\bibfnamefont {Martin}\
  \bibnamefont {Reiris}}, \ and\ \bibinfo {author} {\bibfnamefont {Sergio}\
  \bibnamefont {Dain}},\ }\bibfield  {title} {\enquote {\bibinfo {title}
  {{Black hole Area-Angular momentum inequality in non-vacuum spacetimes}},}\
  }\href {\doibase 10.1103/PhysRevD.84.121503} {\bibfield  {journal} {\bibinfo
  {journal} {Phys.Rev.}\ }\textbf {\bibinfo {volume} {D84}},\ \bibinfo {pages}
  {121503} (\bibinfo {year} {2011}{\natexlab{b}})},\ \Eprint
  {http://arxiv.org/abs/1106.3743} {arXiv:1106.3743 [gr-qc]} \BibitemShut
  {NoStop}%
\bibitem [{\citenamefont {Owen}(2009)}]{Owen:2009sb}%
  \BibitemOpen
  \bibfield  {author} {\bibinfo {author} {\bibfnamefont {Robert}\ \bibnamefont
  {Owen}},\ }\bibfield  {title} {\enquote {\bibinfo {title} {{The Final Remnant
  of Binary Black Hole Mergers: Multipolar Analysis}},}\ }\href {\doibase
  10.1103/PhysRevD.80.084012} {\bibfield  {journal} {\bibinfo  {journal} {Phys.
  Rev.}\ }\textbf {\bibinfo {volume} {D80}},\ \bibinfo {pages} {084012}
  (\bibinfo {year} {2009})},\ \Eprint {http://arxiv.org/abs/0907.0280}
  {arXiv:0907.0280 [gr-qc]} \BibitemShut {NoStop}%
\bibitem [{\citenamefont {Price}(1972)}]{Price:1972pw}%
  \BibitemOpen
  \bibfield  {author} {\bibinfo {author} {\bibfnamefont {Richard~H.}\
  \bibnamefont {Price}},\ }\bibfield  {title} {\enquote {\bibinfo {title}
  {{Nonspherical Perturbations of Relativistic Gravitational Collapse. II.
  Integer-Spin, Zero-Rest-Mass Fields}},}\ }\href {\doibase
  10.1103/PhysRevD.5.2439} {\bibfield  {journal} {\bibinfo  {journal} {Phys.
  Rev.}\ }\textbf {\bibinfo {volume} {D5}},\ \bibinfo {pages} {2439--2454}
  (\bibinfo {year} {1972})}\BibitemShut {NoStop}%
\bibitem [{\citenamefont {Dafermos}\ \emph {et~al.}(2016)\citenamefont
  {Dafermos}, \citenamefont {Holzegel},\ and\ \citenamefont
  {Rodnianski}}]{Dafermos:2016uzj}%
  \BibitemOpen
  \bibfield  {author} {\bibinfo {author} {\bibfnamefont {Mihalis}\ \bibnamefont
  {Dafermos}}, \bibinfo {author} {\bibfnamefont {Gustav}\ \bibnamefont
  {Holzegel}}, \ and\ \bibinfo {author} {\bibfnamefont {Igor}\ \bibnamefont
  {Rodnianski}},\ }\bibfield  {title} {\enquote {\bibinfo {title} {{The linear
  stability of the Schwarzschild solution to gravitational perturbations}},}\
  }\href@noop {} {\  (\bibinfo {year} {2016})},\ \Eprint
  {http://arxiv.org/abs/1601.06467} {arXiv:1601.06467 [gr-qc]} \BibitemShut
  {NoStop}%
\bibitem [{\citenamefont {Kamaretsos}\ \emph {et~al.}(2012)\citenamefont
  {Kamaretsos}, \citenamefont {Hannam}, \citenamefont {Husa},\ and\
  \citenamefont {Sathyaprakash}}]{Kamaretsos:2011um}%
  \BibitemOpen
  \bibfield  {author} {\bibinfo {author} {\bibfnamefont {Ioannis}\ \bibnamefont
  {Kamaretsos}}, \bibinfo {author} {\bibfnamefont {Mark}\ \bibnamefont
  {Hannam}}, \bibinfo {author} {\bibfnamefont {Sascha}\ \bibnamefont {Husa}}, \
  and\ \bibinfo {author} {\bibfnamefont {B.~S.}\ \bibnamefont
  {Sathyaprakash}},\ }\bibfield  {title} {\enquote {\bibinfo {title}
  {{Black-hole hair loss: learning about binary progenitors from ringdown
  signals}},}\ }\href {\doibase 10.1103/PhysRevD.85.024018} {\bibfield
  {journal} {\bibinfo  {journal} {Phys. Rev.}\ }\textbf {\bibinfo {volume}
  {D85}},\ \bibinfo {pages} {024018} (\bibinfo {year} {2012})},\ \Eprint
  {http://arxiv.org/abs/1107.0854} {arXiv:1107.0854 [gr-qc]} \BibitemShut
  {NoStop}%
\bibitem [{\citenamefont {Goldberg}\ \emph {et~al.}(1967)\citenamefont
  {Goldberg}, \citenamefont {MacFarlane}, \citenamefont {Newman}, \citenamefont
  {Rohrlich},\ and\ \citenamefont {Sudarshan}}]{Goldberg:1966uu}%
  \BibitemOpen
  \bibfield  {author} {\bibinfo {author} {\bibfnamefont {J.~N.}\ \bibnamefont
  {Goldberg}}, \bibinfo {author} {\bibfnamefont {A.~J.}\ \bibnamefont
  {MacFarlane}}, \bibinfo {author} {\bibfnamefont {E.~T.}\ \bibnamefont
  {Newman}}, \bibinfo {author} {\bibfnamefont {F.}~\bibnamefont {Rohrlich}}, \
  and\ \bibinfo {author} {\bibfnamefont {E.~C.~G.}\ \bibnamefont {Sudarshan}},\
  }\bibfield  {title} {\enquote {\bibinfo {title} {{Spin s spherical harmonics
  and edth}},}\ }\href@noop {} {\bibfield  {journal} {\bibinfo  {journal} {J.
  Math. Phys.}\ }\textbf {\bibinfo {volume} {8}},\ \bibinfo {pages} {2155}
  (\bibinfo {year} {1967})}\BibitemShut {NoStop}%
\bibitem [{\citenamefont {Gelfand}\ \emph {et~al.}(1963)\citenamefont
  {Gelfand}, \citenamefont {Minlos},\ and\ \citenamefont {Shapiro}}]{GMS}%
  \BibitemOpen
  \bibfield  {author} {\bibinfo {author} {\bibfnamefont {I.~M.}\ \bibnamefont
  {Gelfand}}, \bibinfo {author} {\bibfnamefont {R.~A.}\ \bibnamefont {Minlos}},
  \ and\ \bibinfo {author} {\bibfnamefont {Z.~Ya.}\ \bibnamefont {Shapiro}},\
  }\href@noop {} {\emph {\bibinfo {title} {Representations of the rotation and
  Lorentz groups and their applications}}}\ (\bibinfo  {publisher} {Pergamon
  Press},\ \bibinfo {address} {New York},\ \bibinfo {year} {1963})\BibitemShut
  {NoStop}%
\bibitem [{\citenamefont {Krishnan}(2012)}]{Krishnan:2012bt}%
  \BibitemOpen
  \bibfield  {author} {\bibinfo {author} {\bibfnamefont {Badri}\ \bibnamefont
  {Krishnan}},\ }\bibfield  {title} {\enquote {\bibinfo {title} {{The spacetime
  in the neighborhood of a general isolated black hole}},}\ }\href {\doibase
  10.1088/0264-9381/29/20/205006} {\bibfield  {journal} {\bibinfo  {journal}
  {Class.Quant.Grav.}\ }\textbf {\bibinfo {volume} {29}},\ \bibinfo {pages}
  {205006} (\bibinfo {year} {2012})},\ \Eprint {http://arxiv.org/abs/1204.4345}
  {arXiv:1204.4345 [gr-qc]} \BibitemShut {NoStop}%
\bibitem [{\citenamefont {Lewandowski}(2000)}]{Lewandowski:1999zs}%
  \BibitemOpen
  \bibfield  {author} {\bibinfo {author} {\bibfnamefont {Jerzy}\ \bibnamefont
  {Lewandowski}},\ }\bibfield  {title} {\enquote {\bibinfo {title} {{Spacetimes
  Admitting Isolated Horizons}},}\ }\href {\doibase 10.1088/0264-9381/17/4/101}
  {\bibfield  {journal} {\bibinfo  {journal} {Class. Quant. Grav.}\ }\textbf
  {\bibinfo {volume} {17}},\ \bibinfo {pages} {L53--L59} (\bibinfo {year}
  {2000})},\ \Eprint {http://arxiv.org/abs/gr-qc/9907058} {arXiv:gr-qc/9907058}
  \BibitemShut {NoStop}%
\bibitem [{\citenamefont {Lewandowski}\ and\ \citenamefont
  {Paw{\l}owski}(2014)}]{Lewandowski:2014nta}%
  \BibitemOpen
  \bibfield  {author} {\bibinfo {author} {\bibfnamefont {Jerzy}\ \bibnamefont
  {Lewandowski}}\ and\ \bibinfo {author} {\bibfnamefont {Tomasz}\ \bibnamefont
  {Paw{\l}owski}},\ }\bibfield  {title} {\enquote {\bibinfo {title}
  {{Neighborhoods of isolated horizons and their stationarity}},}\ }\href
  {\doibase 10.1088/0264-9381/31/17/175012} {\bibfield  {journal} {\bibinfo
  {journal} {Class. Quant. Grav.}\ }\textbf {\bibinfo {volume} {31}},\ \bibinfo
  {pages} {175012} (\bibinfo {year} {2014})},\ \Eprint
  {http://arxiv.org/abs/1404.7836} {arXiv:1404.7836 [gr-qc]} \BibitemShut
  {NoStop}%
\bibitem [{\citenamefont {Scholtz}\ \emph {et~al.}(2017)\citenamefont
  {Scholtz}, \citenamefont {Flandera},\ and\ \citenamefont
  {Guerlebeck}}]{Scholtz:2017ttf}%
  \BibitemOpen
  \bibfield  {author} {\bibinfo {author} {\bibfnamefont {Martin}\ \bibnamefont
  {Scholtz}}, \bibinfo {author} {\bibfnamefont {Ales}\ \bibnamefont
  {Flandera}}, \ and\ \bibinfo {author} {\bibfnamefont {Norman}\ \bibnamefont
  {Guerlebeck}},\ }\bibfield  {title} {\enquote {\bibinfo {title} {{Kerr-Newman
  black hole in the formalism of isolated horizons}},}\ }\href {\doibase
  10.1103/PhysRevD.96.064024} {\bibfield  {journal} {\bibinfo  {journal} {Phys.
  Rev.}\ }\textbf {\bibinfo {volume} {D96}},\ \bibinfo {pages} {064024}
  (\bibinfo {year} {2017})},\ \Eprint {http://arxiv.org/abs/1708.06383}
  {arXiv:1708.06383 [gr-qc]} \BibitemShut {NoStop}%
\bibitem [{\citenamefont {Guerlebeck}\ and\ \citenamefont
  {Scholtz}(2018)}]{Gurlebeck:2018smy}%
  \BibitemOpen
  \bibfield  {author} {\bibinfo {author} {\bibfnamefont {Norman}\ \bibnamefont
  {Guerlebeck}}\ and\ \bibinfo {author} {\bibfnamefont {Martin}\ \bibnamefont
  {Scholtz}},\ }\bibfield  {title} {\enquote {\bibinfo {title} {{The Meissner
  Effect for axially symmetric charged black holes}},}\ }\href@noop {} {\
  (\bibinfo {year} {2018})},\ \Eprint {http://arxiv.org/abs/1802.05423}
  {arXiv:1802.05423 [gr-qc]} \BibitemShut {NoStop}%
\bibitem [{\citenamefont {Bartnik}\ and\ \citenamefont
  {Isenberg}(2006)}]{Bartnik:2005qj}%
  \BibitemOpen
  \bibfield  {author} {\bibinfo {author} {\bibfnamefont {Robert}\ \bibnamefont
  {Bartnik}}\ and\ \bibinfo {author} {\bibfnamefont {James}\ \bibnamefont
  {Isenberg}},\ }\bibfield  {title} {\enquote {\bibinfo {title} {{Spherically
  symmetric dynamical horizons}},}\ }\href {\doibase
  10.1088/0264-9381/23/7/020} {\bibfield  {journal} {\bibinfo  {journal}
  {Class. Quant. Grav.}\ }\textbf {\bibinfo {volume} {23}},\ \bibinfo {pages}
  {2559--2570} (\bibinfo {year} {2006})},\ \Eprint
  {http://arxiv.org/abs/gr-qc/0512091} {arXiv:gr-qc/0512091} \BibitemShut
  {NoStop}%
\bibitem [{\citenamefont {Chandrasekhar}(1985)}]{Chandrasekhar:1985kt}%
  \BibitemOpen
  \bibfield  {author} {\bibinfo {author} {\bibfnamefont {S.}~\bibnamefont
  {Chandrasekhar}},\ }\href@noop {} {\emph {\bibinfo {title} {{The mathematical
  theory of black holes}}}}\ (\bibinfo  {publisher} {Oxford Classic Texts in
  the Physical Sciences},\ \bibinfo {year} {1985})\BibitemShut {NoStop}%
\end{thebibliography}%

\end{document}